\documentclass{aa}
\usepackage{epsfig}
\usepackage{amssymb}
 \def\mso{\,\mathrm{M}_\odot}                                                  
 \def\rso{\,{\rm R}_\odot}                                                      
 \def\lso{\,{\rm L}_\odot}                                                      

 \def\kms{\, {\rm km}\, {\rm s}^{-1}}                                           
 \def\egs{\, {\rm erg}\, {\rm s}^{-1}}

 \def\h1{\hangindent=1.0truecm \hangafter=0}                                    
 \def\simle{\mathrel{\hbox{\rlap{\hbox{\lower4pt\hbox{$\sim$}}}\hbox{$<$}}}}    
 \def\simgr{\mathrel{\hbox{\rlap{\hbox{\lower4pt\hbox{$\sim$}}}\hbox{$>$}}}}    
 \def\msoy{\, \mso~{\rm yr}^{-1}}

 \def\hb{\hfill\break}
 \def\h2{\hb\noindent \hangindent=0.5cm \hangafter=1}

 \def\utw{\smash{\rlap{\lower5pt\hbox{$\sim$}}}}
 \def\udtw{\smash{\rlap{\lower6pt\hbox{$\approx$}}}}

\begin{document}
\thesaurus{ 06 (08.01.1; 08.02.1; 08.05.3; 08.09.3; 08.19.4; 13.25.5)}

\title{The evolution of main sequence star + white dwarf binary
systems towards Type~Ia supernovae }

\author{N. Langer\inst{1,2}, A. Deutschmann\inst{2}, S. Wellstein\inst{2} 
  \and P. H\"oflich\inst{3} }

\institute{
Astronomical Institute, Utrecht University, Princetonplein 5,
NL-3584 CC, Utrecht, The Netherlands
\and
Institut f\"ur Physik, Universit\"at
Potsdam, Am Neuen Palais 10, D--14415~Potsdam, Germany
\and
Department of Astronomy, University of Texas, Austin, TX 78712, USA 
}

\offprints { N. Langer (email: {\tt N.Langer@ astro.uu.nl})}

\date{Received  ; accepted , }

\authorrunning { N. Langer et al.}
\titlerunning {Main sequence star + white dwarf binaries as SN~Ia progenitors}

\maketitle

\begin{abstract}

Close binaries consisting of a main sequence star and a white dwarf are
considered as candidates for Type~Ia supernova progenitors. 
We present selfconsistent calculations of 
the time dependence of the structure of the main sequence star,
the mass transfer rate, and the orbit
by means of a binary stellar evolution program.
We obtain, for the first time, a complete picture of the time
evolution of the mass transfer rate in such systems.
We find a long switch-on phase of the mass transfer,
about $10^6\,$yr, during which nova outbursts should persist in all
systems. Furthermore, we find that the white dwarfs can reach
the Chandrasekhar mass only during the decline phase of the mass transfer,
which may have consequences for the critical accretion rate
for stationary nuclear burning on the white dwarf surface.
In contrast to results based on simple estimates of the mass transfer rate 
in systems of the considered type, our results allow for the
possibility that even systems with rather small initial white dwarf masses
($\sim 0.7\mso$) may produce Type~Ia supernovae, which then might
originate from very rapidly rotating white dwarfs.

We present results for two different metallicities, Z=0.02 and Z=0.001.
We find that for systems with the lower metallicity, the mass transfer rates are
on average five times larger than in comparable system at solar metallicity.
This leads to a systematic shift of the supernova~Ia progenitor population.
Firstly, while for Z=0.02 --- for our choice of white dwarf wind mass loss and
mass accumulation rate --- 
donor star initial masses in supernova progenitor systems are restricted to 
the range 1.6$\mso$...2.3$\mso$, they are in the interval
1.4$\mso$...1.8$\mso$ at low~Z. Secondly, the initial white dwarf masses
need, on average, to be larger by 0.2$\mso$ at low~Z in order to obtain
a Chandrasekhar mass white dwarf. This metallicity dependences 
have very little effect on the progenitor life times, but may be
responsible for a drop of the Type~Ia supernova rate for low
metallicity, and may introduce a Z-dependence in the properties of
supernovae which stem from close main sequence star~+~ white dwarf systems.

We estimate the X-ray luminosities of the computed systems,
and investigate their donor star and orbital properties.
We find the donor stars to be underluminous by up to one order of
magnitude, and more compact than normal main sequence stars.
In general, our systems correspond well to observed close binary
supersoft X-ray sources. We further derive the chemical and kinematical
properties of the stellar remnants of our systems after the explosion
of the white dwarf, which may serve as a test of the viability of
the considered Type~Ia supernova scenario.


\end{abstract}

\section{Introduction}


During the last years, the refinement of supernova observations, 
e.g., the routine detection of supernovae at large redshifts,
has made them a powerful tool to probe cosmology.  
It allowed to determine the Hubble constant with
unprecedented accuracy (Riess et al. 1995, Hamuy et al. 1996; see also
H\"oflich \& Khokhlov 1996). Even more exciting,
recent results (e.g., Riess et al. 1999, Perlmutter et al. 1999)
are consistent with a low matter density in the Universe and, 
intriguingly, hints for a positive cosmological constant.  
These findings are based on empirical brightness-decline relations which are
calibrated locally. 
This leaves potential systematic effects of 
supernova~Ia properties with redshift
as major concern. To this end, it would be desirable to
obtain an estimate of such effects from theoretical models of 
supernova~Ia progenitor systems.

However, despite considerable efforts during the last decades, the exact 
nature of supernova~Ia progenitors is still unclear.
On observational and theoretical grounds, it is generally agreed that 
Type~Ia supernovae result from the thermonuclear disruption of a CO white dwarf
(e.g., Woosley \& Weaver 1986, Wheeler 1996, Nomoto et al. 1997,
Branch 1998). Since isolated white dwarfs cool, a close binary component
which transfers mass to the
white dwarf is a prerequisite to obtain a Type~Ia supernova. 
Various binary evolution
scenarios leading to exploding CO~white dwarfs 
have been proposed and investigated,
but hitherto it is unclear
which of them is preferred in nature (cf. Branch 1988, Livio 1999).

In this paper, we study the evolution of close binary systems
consisting of a carbon-oxygen white dwarf 
and a main sequence star, which was repeatedly
proposed as promising supernova~Ia progenitor scenario
(cf. Nomoto \& Sugimoto 1977, Li \& van den Heuvel
1997, Kato \& Hachisu 1999, Hachisu et al. 1999). In these systems,
the carbon-oxygen white dwarfs are the remainders of stars
with an initial mass below $\approx 10\mso$ which have lost
their H/He-rich envelope, with CO cores of $\approx 0.6...1.2\mso$.
If accretion is sufficiently fast the accreted hydrogen may burn
to helium and, subsequently, to CO on the surface of the white dwarf, 
and its mass
grows close to the Chandrasekhar mass.

The binary evolution leading to close white dwarf~+~main sequence star 
systems is not yet
well understood (cf., Livio 1996). However, we know a large number
($\sim 10^3$) of close white dwarf~+~main sequence star 
systems as Cataclysmic Variables
(Ritter \& Kolb 1998), most of which do not evolve into Type~Ia supernovae
since they undergo nova outbursts which may prevent a secular increase
of the white dwarf mass (e.g., Kovetz \& Prialnik 1997).
The idea that also the slightly more massive  
systems of the same type studied here
occur in nature is supported by
population synthesis studies, which predict their birth rate to
be comparable, within an order of magnitude, to the observed rate
of Ia~supernovae (e.g., de~Kool \& Ritter 1993, Rappaport et al, 1994).
It is further supported through the discovery of the so 
called supersoft X-ray sources (Greiner et al. 1991, 
Kahabka \& van den Heuvel 1997), which may represent the
observational counterparts of the binary systems studied here
theoretically.

We investigate the properties of close main sequence star-white dwarf
systems at two different metallicities.
As we derive the detailed time-dependence of the accretion rate,
our work is relevant for the understanding of individual
supernovae and supersoft X-ray binaries, 
for the change of their average properties with
metallicity, and for the dependence of the rate of Ia~supernovae 
with metallicity.
We introduce our computational method in Sect.~2, and present
our results for the mass transfer rate and resulting maximum white dwarf
masses in Sect.~3. In Sect.~4, we discuss the evolution of
the white dwarf spin, of the binary orbit and of the main sequence
stars. In Sect.~5, we compare our results with observations of
supersoft X-ray sources and derive clues which may help to 
identify the remaining main sequence 
star in a supernova~Ia remnant. Our conclusions are given in Sect.~6.

\section{Computational method and physical assumptions}

The numerical models presented in this work are computed with a binary
stellar evolution code developed by Braun (1997) on the basis of a single
star code (Langer 1998, and references therein). It is a 1-dimensional
implicit Lagrangian code which solves the hydrodynamic form of the
stellar structure and evolution equations (Kippenhahn \& Weigert 1990).
The evolution of two stars and, in case of mass transfer, the evolution
of the mass transfer rate and of the orbital separation are computed
simultaneously through an implicit coupling scheme (see also 
Wellstein \& Langer 1999, Wellstein et al. 1999), using the 
Roche-approximation in the formulation of Eggleton (1983).
To compute the mass transfer rate, we use the description of Ritter (1988).
The stellar models are computed using OPAL opacities (Iglesias \& Rogers
1996) and extended nuclear networks including the pp~I,~II, and~III chains,
the four CNO-cycles, and the NeNa- and MgAl-cycles (cf. Arnould et al. 1999). 

\subsection{The mass accretion rate of the white dwarf}

In order to compute the evolution of close main sequence star-white dwarf
pairs, we invoke the following assumptions. In the computation of the
binary system the white dwarf is approximated as a point mass
(however, cf. Sect.~2.3), while the main sequence star is resolved
with typically 1000 grid points. The systems are started at time
t:=0 with a zero age main sequence star of mass $M_{\rm 1,i}$ and an
arbitrary white dwarf mass $M_{\rm WD,i}$ at an arbitrary orbital separation 
$d_{\rm i}$. For most models, we use $M_{\rm WD,i}=0.8\mso$ or $1.0\mso$.
For $d_{\rm i}$ we consider only values which lead to mass transfer during the
core hydrogen burning phase of the main sequence star, i.e. so called
Case~A mass transfer.

The mass of the white dwarf (the point mass) is allowed to vary in
accordance with critical mass transfer rates which were taken from the
literature as follows. For mass transfer rates $\dot M \geq
\dot M_{\rm H}(M_{\rm WD})$ and
$\dot M \geq \dot M_{\rm He}(M_{\rm WD})$,
we allow the white dwarf mass to increase. Here, $\dot M_{\rm H}$ and
$\dot M_{\rm He}$ are the critical accretion rates above which 
H- or He-burning proceeds such that violent nova flashes and consequent
mass ejection from the white dwarf are avoided. 
We adopt $\dot M_{\rm He} = 10^{-8}\msoy$ for models with a metallicity
of $Z=2$\%, and $4\, 10^{-8}\msoy$ for $Z=0.001$
(Fujimoto 1982, Nomoto \& Kondo 1991).
For $\dot M_{\rm H}$ we rely on Figure~5 of Kahabka \& van den Heuvel (1997).
For $\dot M_{\rm He} > \dot M > \dot M_{\rm H}$ we assume the
white dwarf mass to grow as well, but by accumulating a degenerate thick helium
layer. For $\dot M \le \dot M_{\rm Edd} := L_{\rm Edd} / \varepsilon$
and $\dot M \le \dot M_{\rm RG}$, 
we assume $\dot M_{\rm WD} = \dot M$. Here, 
$M_{\rm RG}$ is the critical accretion rate above which the white dwarf
is assumed to expand to red giant dimensions (Nomoto \& Kondo 1991),
and
\begin{equation}
L_{\rm Edd} = {4\pi c G M_{\rm WD} \over 0.2*(1+X)}
\end{equation}
is the Eddington luminosity of the white dwarf, using $0.2*(1+X)$ as the
opacity coefficient due to electron scattering with $X$ being the
hydrogen mass fraction. The quantity 
$\varepsilon = 7\, 10^{18}\,$erg$\,$g$^{-1}$
gives the approximate amount of energy obtained per gram of hydrogen
burnt into helium or carbon/oxygen. 
For larger mass transfer rates we assume that
the white dwarf has a wind which carries the excess mass away
(Hachisu et al. 1996). We stop our calculations for models with
$\dot M_{\rm wind} > 3 \dot M_{\rm Edd}$.

Our assumptions concerning the critical accretion rates are similar to those
of Li \& van den Heuvel (1997).
However, we deviate from them by adopting a
maximum possible wind mass loss rate of 
$\dot M_{\rm wind} > 3 \dot M_{\rm Edd}$.
For $\dot M = 2 \dot M_{\rm Edd}$, the wind momentum $\dot M_{\rm wind}
v_{\infty} = \dot M_{\rm Edd} v_{\infty}$ is of the order of the photon 
momentum $L_{\rm Edd}/c$. 
More specifically, $\dot M_{\rm wind} = \dot M_{\rm Edd}$ implies
\begin{equation}
{\dot M_{\rm wind} v_{\infty} \over L/c} = {v_{\infty} c \over
\varepsilon}
\end{equation}
which, for
\begin{equation}
v_{\infty} \simeq v_{\rm escape} = \sqrt{2G M_{\rm WD}\over R_{\rm WD}},
\end{equation}
is of order unity.
Our restriction implies that the winds we
invoke remain in a regime where the wind efficiency is undisputed 
(cf. Lamers \& Cassinelli 1999).
It limits our mass loss to rates 
well below those allowed by Li \& van den Heuvel (1997).
I.e., with $\dot M_{\rm Edd} \simeq 3.3\, 10^{-7} (M_{\rm WD}/\mso ) \msoy$,
our upper limit is of the order of $10^{-6}\msoy$ rather than $10^{-4}\msoy$.

In this context we note that there is also an energy limit to
radiation driven winds such that 
$\dot M < \dot M_{\rm En} := L / v_{\infty }^2$; i.e.
\begin{equation}
\dot M_{\rm En} = {L \over L_{\rm Edd}} \left({v_{\rm escape}\over
v_{\infty}}\right)^2 ~ {2\pi c R_{\rm WD}\over 0.2*(1+X)}  .
\end{equation}
This assumes spherical symmetry and ignores the thermal energy
of the wind. It implies that
a star with $\dot M = \dot M_{\rm En}$ is invisible, as {\em all} the
photon energy is used to drive the wind.
With $L=L_{\rm Edd}$ and $v_{\infty} \simeq v_{\rm escape}$, this
results in $\dot M_{\rm En} = 6\, 10^{-6} \left( R_{\rm WD}/ 0.01\rso 
\right) \msoy$. 

Kato \& Iben (1992) and Kato \& Hachisu (1994) have worked
out a theory for optically thick winds which allows to obtain
mass loss rates which can carry well above 100 times the photon
momentum. This has been used by Hachisu et al. (1996) and Li \& van den
Heuvel (1997).  However, such optically thick winds from stars near the
Eddington limit may involve processes which limit the wind efficiency.
For example, if the underlying star carries noticeable amounts of angular
momentum it may reach critical rotation {\em before} reaching the
Eddington limit (Langer 1997, 1998). This means that only a fraction of
the stellar surface, around the equator, is experiencing the critical outflow
condition, rather than the whole stellar surface. The reality of this 
phenomenon is demonstrated by the highly non-spherical, axially
symmetric nebulae around Luminous Blue Variables (Nota et al. 1995),
whose outbursts are likely driven by super-Eddington winds (cf. Langer
2000). Also nova winds and outflows from supersoft X-ray sources
are known to be highly anisotropic. Another limiting factor 
is convection or turbulent energy transport, which can be very
important in Eddington flows (Heger \& Langer 1996,
Owocki \& Gayley 1997, Langer 1997).
Furthermore, authors who compute the driving force in winds of
Wolf-Rayet stars using detailed non-equilibrium atomic physics
in order to compute the photon trapping in the optically thick parts
of the wind flow, rather than relying on the continuum approximation
\footnote{We note in passing that in order to derive the
driving force from the continuum approximation,
the flux-mean opacity coefficient needs to be used,
not the Rosseland mean opacity, which is often used instead.},
find that only about 5\% of the stellar photon luminosity
is converted into kinetic wind energy (Lucy \& Abbott 1993). 
Were this number valid for white dwarfs at the Eddington limit, 
it would imply a mass loss rate of 
$\dot M = 0.05 \dot M_{\rm E}  = 
3\, 10^{-7} \left( R_{\rm WD}/ 0.01\rso \right) \msoy$.   


The consequences of our conservative assumption
on the white dwarf wind efficiency, which deviates from assumptions
on wind mass loss rates in previous studies (cf. Hachisu et al. 1996,
Kobayashi et al. 1998), are discussed in Sect.~3.2.
Self-excited wind as proposed by King \& van Teeseling (1998) are not
considered here.

Other than Li \& van den Heuvel (1997), we do not allow for 
the possibility of the partial mass ejection in case of weak shell
flashes, as we think that the current uncertainties 
(cf. Prialnik \& Kovetz 1995) may not make
such sophistication worthwhile but rather complicate the understanding
of the obtained results. 
Our value of $\dot M_{\rm H} = 10^{-8}\msoy$ for $Z=2$\%
is in agreement with the general conclusion of Prialnik \& Kovetz that
$\dot M \simgr 10^{-8}\msoy$ leads to growing white dwarf masses. 
We also do not include a reduction of the mass
accumulation efficiency of the white dwarf due to winds excited
by helium shell flashes (Kato \& Hachisu 1999).
However, in order to study the influence of the threshold value for mass
accumulation on the white dwarf, 
we investigate the effect of an increase of this 
value by one order of magnitude, as outlined in Sect.~3.2.

\subsection{Further white dwarf properties}

Basic properties of the white dwarf are estimated as follows:
Its  radius $R_{\rm WD}$ is given by the mass-radius relation 
\begin{equation}
R_{\rm WD} = f M_{\rm WD}^{-1/3} 
\end{equation}
with
$f={2\over G} ({3\over 8\pi})^{4/3} h^2 / 
\left(2^{1/3} \mu_e^{5/3} m_p^{5/3} m_e\right)
\simeq 9.03\, 10^{19}\,$cm$\,$g$^{1/3}$ (Kippenhahn \& Weigert 1990;
see also Nauenberg 1972, Provencal et al. 1998).

We assume that the radiation from the white dwarf can be 
approximated by a Black Body with an
effective temperature 
\begin{equation}
T_{\rm WD}^4 = {\varepsilon \dot M_{\rm WD} \over 4\pi R_{\rm WD}^2 \sigma} ,
\end{equation}
with $\sigma = 5.67\, 10^{-5}\,$erg$\,$s$^{-1}\,$cm$^{-2}\,$K$^{-4}$ 
being the Stefan-Boltzmann constant. 

We estimate the white dwarf's angular momentum and rotation
frequency by assuming it to be zero initially, and that the specific
angular momentum of the accreted matter is 
$j = v_{\rm Kepler} R_{\rm WD}$, with
$v_{\rm Kepler} = \sqrt{G M_{\rm WD} / R_{\rm WD}}$.
The equation for the accumulated angular momentum
\begin{equation}
J := \int_{t^{\prime}=0}^t  \dot M_{\rm WD} j dt^{\prime}  ,
\end{equation}
yields
\begin{equation}
J = {3\over 4} \sqrt{Gf} \left( M_{\rm WD}^{4/3} - M_{\rm WD,i}^{4/3}
\right) . 
\end{equation}
Assuming then rigid rotation for the white dwarf interior
(cf. Kippenhahn 1974) we can estimate its angular velocity as
\begin{equation}
\omega = {J \over k^2 M_{\rm WD} R_{\rm WD}^2}
\end{equation}
where $k$ is the dimensionless radius of gyration .
The ratio $\Omega = \omega / \omega_{\rm Kepler}$ becomes
\begin{equation}
\Omega = {3\over 4k^2} \left(1-\left({M_{\rm WD,i}\over M_{\rm WD}}
\right)^{4/3}\right)  
\end{equation}
(cf. Papaloizou \& Pringle 1978, Livio \& Pringle 1998).
The implications of this relation are discussed in Sect.~4.1.

\subsection{Single star evolution}

We have constructed stellar model sequences for two metallicities,
$Z=0.02$ and $Z=0.001$. The initial helium mass fraction is computed as
$Y=Y_{\rm pm} + (dY/dZ) Z$, using $Y_{\rm pm}=0.24$ as 
primordial helium mass fraction and $dY/dZ=2$. 
The resulting values are $Y=0.280$ and $Y=0.242$ for the high
and low metallicity considered here, respectively.
The relative abundances of the metals are chosen according
to the solar system abundances (Grevesse \& Sauval 1998).
We have computed all models with extended convective cores 
(``overshooting'') by 0.2~pressure scale heights 
($\alpha_{\rm over}=0.2$). The resulting tracks in the HR diagram 
are very similar to those of Schaller et al. (1992), that of 
our $Z=0.02$ models for 2$\mso$ and 1.7$\mso$ and of our
2$\mso$ sequence at $Z=0.001$ --- the only three sequences which
we can directly compare --- are virtually identical. 

Figs.~1 and~2 show the time evolution of the radii during
core hydrogen burning for our single star models
at the two metallicities considered in this work,  
for the mass range which is relevant
in the context of this paper. During core hydrogen burning, the radii
increase by factors 
2.0...2.5 and 1.7...2.3 for the higher and lower considered metallicity,
respectively, with larger values corresponding to larger masses.

\begin{figure}[t]
\begin{centering}
\epsfxsize=0.9\hsize
\epsffile{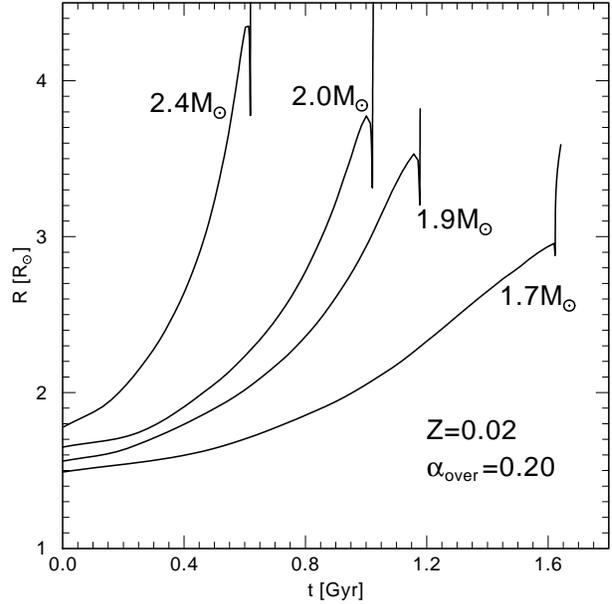}
\caption{Evolution of the stellar radius as a function of time
for single stars of about solar metallicity (Z=0.02) in the mass range 
from 1.7 to 2.4$\mso$ computed with convective core overshooting
($\alpha_{\rm over}=0.2$), from the zero age main sequence until shortly
after core hydrogen exhaustion.
}
\end{centering}
\end{figure}

\begin{figure}[ht]
\begin{centering}
\epsfxsize=0.9\hsize
\epsffile{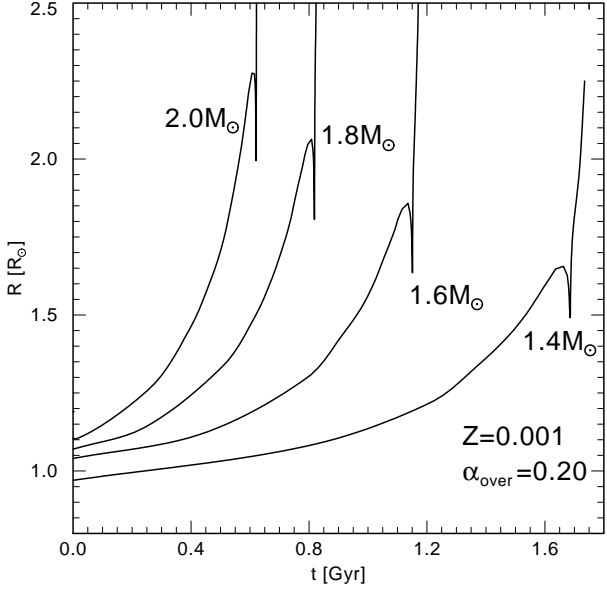}
\caption{Same as Figure~1, but for low metallicity stars (Z=0.001) in
the mass range 1.4 to 2$\mso$.
}
\end{centering}
\end{figure}

\begin{table}
\begin{center}
\caption{Comparison of surface properties and of the
Kelvin-Helmholtz time scale $\tau_{\rm KH,TAMS} = G M^2/(2 R L)$
for models of $2\mso$ sequences computed with
metallicities of $Z=0.02$ and $Z=0.001$, at the zero age main
sequence (ZAMS) and the terminal age main sequence (TAMS)}
\vbox{
\begin{tabular}{l c c}
\hline
~ & $Z=0.02$ & $Z=0.001$ \\
\hline
$L_{\rm ZAMS}$        & 15.9$\lso$        & 26.3$\lso$       \\
$T_{\rm eff, ZAMS}$   & 9158 K            & 12474 K          \\
$R_{\rm ZAMS}$        & 1.60$\rso$        & 1.11$\rso$       \\
$\tau_{\rm KH,ZAMS}$  & $2.5$Myr          & $2.2$Myr         \\
\vspace{-3mm}&&\\
$L_{\rm TAMS}$        & 22.3$\lso$        & 56.1$\lso$       \\
$T_{\rm eff, TAMS}$   & 6600 K            & 10452 K          \\
$R_{\rm TAMS}$        & 3.65$\rso$        & 2.31$\rso$       \\
$\tau_{\rm KH,TAMS}$  & $0.75$Myr         & $0.48$Myr        \\
\hline
\end{tabular}
}
\end{center}
\end{table}

The radii of the metal poor stars are nearly a factor of~2 smaller
than those of stars with a comparable mass and evolutionary stage at $Z=0.02$.
This has consequences for the binary evolution models discussed
below, i.e.,  the orbital periods in Case~A systems are much smaller for
smaller metallicity.  Furthermore, the metal poor main sequence stars
are hotter and more luminous compared to stars of the same mass at
$Z=0.02$. This is in agreement with previous models of stars of
comparable masses and metallicities (e.g., Schaller et al. 1992).
Table~1 gives the quantitative details of models from our 2$\mso$ 
sequences at $Z=0.02$ and $Z=0.001$ at the beginning and at the end
of core hydrogen burning. Although the models at lower
metallicities are more compact, they are also more luminous than
the metal richer models, with the consequence of a shorter 
Kelvin-Helmholtz time scale. This has consequences for the mass transfer
rates, as discussed in the next section.

\subsection{Binary evolution: examples}

\begin{figure}[t]
\begin{centering}
\epsfxsize=0.9\hsize
\epsffile{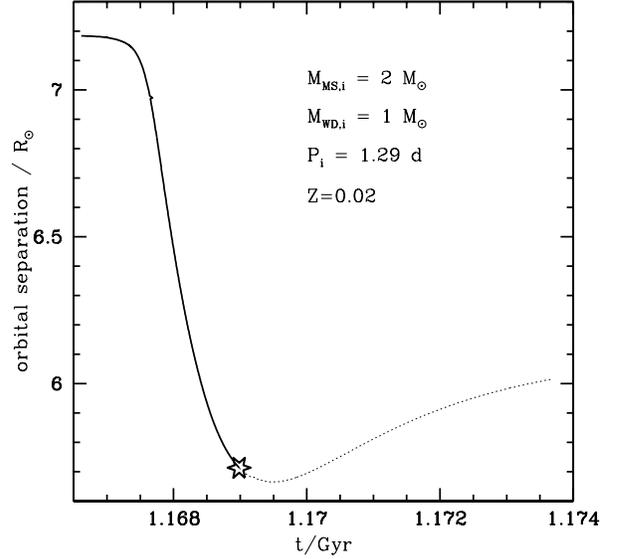}
\caption{Evolution of the orbital period as function of time
for System~No.~6 (cf. Table~1). The curve starts at the onset of mass
transfer at $t\simeq 1.166\,$Gyr. At $t\simeq 1.169$, the white dwarf
mass has grown to 1.4$\mso$; this time is marked by an asterisk on the 
curve. The dashed part of the curve shows
the continuation of the orbital period evolution assuming 
that the white dwarf does not perform a supernova explosion.
}
\end{centering}
\end{figure}

\begin{figure}[t]
\begin{centering}
\epsfxsize=0.9\hsize
\epsffile{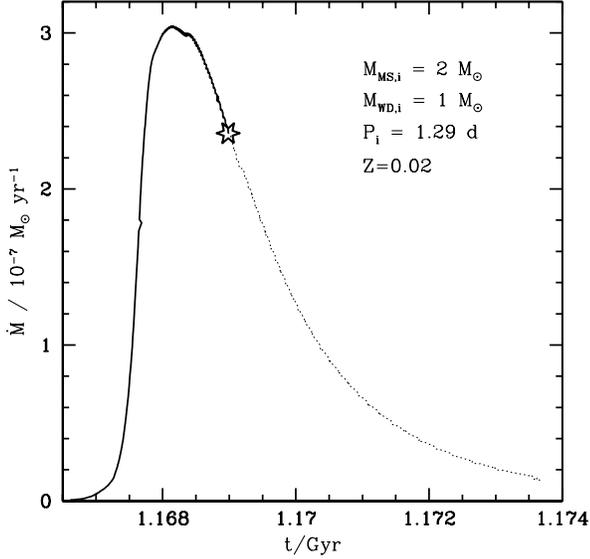}
\caption{Evolution of the mass transfer rate as function of time
for System~No.~6 (cf. Table~1). The curve starts at the onset of mass
transfer at $t\simeq 1.166\,$Gyr. At $t\simeq 1.169$, the white dwarf
mass has grown to 1.4$\mso$; this time is marked by an asterisk on the
curve. The dashed part of the curve shows
the continuation of the mass transfer rate evolution assuming
that the white dwarf does not perform a supernova explosion.
}
\end{centering}
\end{figure}

To illustrate our approach, we consider our System No.~6. Initially, it
consists of a 2$\mso$ zero age main sequence star and a 1$\mso$ white
dwarf (treated as point mass) in a circular orbit with a separation
of $d_{\rm i} = 7.19\rso$. According to Kepler's third law, the initial
period $P_{\rm i}$ is
\begin{equation}
P_{\rm i} = { 2\pi \over \sqrt{G}}
   \sqrt{ {d_{\rm i}^3 \over  M_{\rm MS,i} + M_{\rm WD,i}} }
\end{equation}
i.e., $P_{\rm i} = 1.29\,$d in this case. With this initial set-up,
we skip the previous evolution of the system, i.e. the evolution of the
white dwarf progenitor and the common envelope and spiral-in phase which
brought the two stars close together. 

We set the time $t=0$ at core
hydrogen ignition of our main sequence star. 
In principle, this neglects the duration of the previous evolution
of the system, i.e. the evolutionary time of the white dwarf progenitor.
However, as we shall see
below we deal with rather high initial white dwarf masses, i.e.
relatively massive white dwarf progenitor stars ($\simgr 5\mso$). Therefore,
as the white dwarf starts to accrete at a system age of the order of
the evolutionary time scale of the donor star (1.4...2.4$\mso$; see below)
the so defined time yields a good estimate for the age of the systems
at the time of the supernova explosion (cf. Umeda et al. 1999).

In our System~No.~6, mass transfer starts at $t\simeq 1.166\,$Gyr,
at a central hydrogen mass fraction of the main sequence star of
$X_{\rm c}\simeq 0.17$. The radius of the star has then grown from
$R_{\rm i}\simeq 1.60\rso$ at the ZAMS to 
$R\simeq 3.16\rso$, as we use Eggleton's (1983) approximation
for the Roche radius of the main sequence star
\begin{equation}
R_{\rm L} = d {0.49 q^{2/3} \over 0.6 q^{2/3} \ln (1+q^{1/3}) }
\end{equation}
with $q:=M_{\rm MS}/M_{\rm WD}$.

During the initial phase of the mass transfer evolution, the mass 
transfer rate $\dot M$ is still smaller than $\dot M_{\rm H}$
(cf. Sect.~2.1), and the white dwarf mass can not increase.
Instead, all accreted mass is assumed to be lost in nova outbursts,
carrying away the specific orbital angular momentum of the white dwarf,
leading to a decrease of the orbital separation (cf. Podsiadlowski 
et al. 1992). The amount of mass lost during this phase, and consequently
the change of the orbital parameters, is quite insignificant in
most cases. E.g., in our example $0.003\mso$ are transferred and lost
in nova outbursts. However, we emphasise that the time scale of this 
first phase may be non-negligible, as it may be of the same order of
magnitude as the major mass accretion phase (cf. Sect.~3).

When the mass transfer rate exceeds the critical rates for hydrogen and
helium burning (Sect.~2.1), we allow the white dwarf mass to grow.
As angular momentum conservation again leads to a shrinkage of the
orbit as long as $M_{\rm MS} > M_{\rm WD}$, the mass transfer is thermally
unstable. Since the mass-radius exponents of our main sequence stars are
positive, i.e., mass loss leads to smaller radii (Ritter 1996),
the mass transfer is stabilised due to the thermal disequilibrium
of the main sequence star, and the resulting mass transfer rates
are of the order of magnitude of
\begin{equation}
\dot M \simeq (M_{\rm MS,i} - M_{\rm WD,i})/ \tau_{\rm KH}
\end{equation}
(e.g., Rappaport et al. 1994; however, see Sect.~3.1). 

\begin{figure}[t]
\begin{centering}
\epsfxsize=0.9\hsize
\epsffile{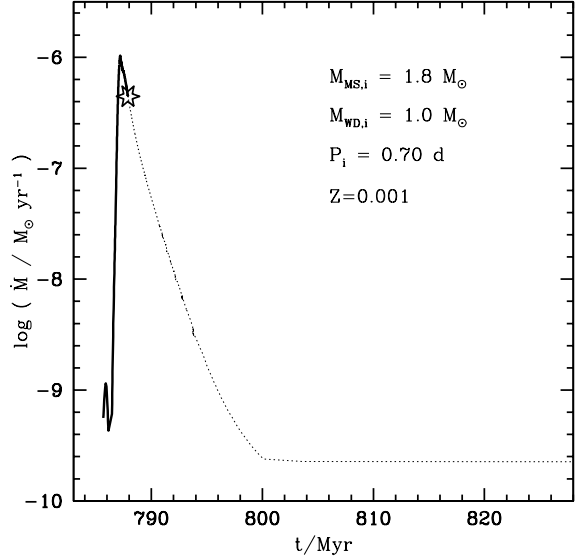}
\caption{Evolution of the mass transfer rate as function of time
for System No.~61 (cf. Fig.~8).}
\end{centering}
\end{figure}


\begin{figure}[ht]
\begin{centering}
\epsfxsize=0.9\hsize
\epsffile{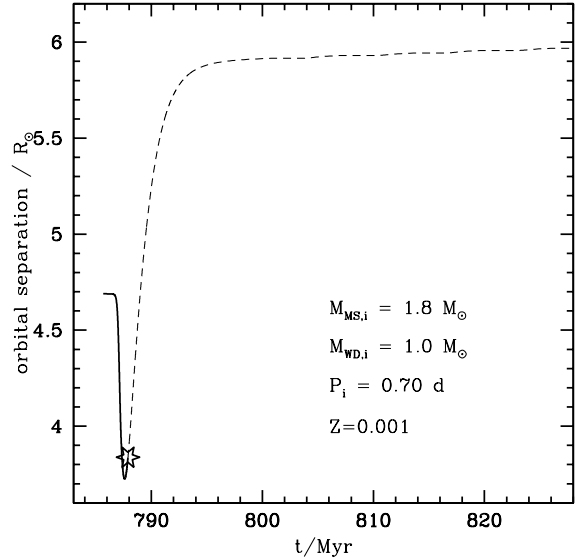}
\caption{Evolution of the orbital separation as function of time
for System No.~61.}
\end{centering}
\end{figure}

Fig.~3 shows the evolution of the orbital separation with time for
System No.~6 throughout the phase of the thermally unstable mass transfer.
Within our approximations, the white dwarf mass reaches $1.4\mso$ at
$t\simeq 1.169\,$Gyr. Although it is likely that the white dwarf would explode
roughly at this point (see below), we follow the further evolution of the
system ignoring this, for several reasons. Most important, the continued
evolution allows us to estimate how changes in our basic assumptions
might affect the fate of the white dwarf. E.g., since it appears to be
undisputed that the white dwarf mass can grow for mass transfer rates
$\dot M \simgr 10^{-7}\msoy$ (Nomoto \& Kondo 1991, Prialnik \& Kovetz 1995,
Kahabka \& van den Heuvel 1997), we define $M_7 := \int \dot M_7 dt$ with
\begin{equation}
\dot M_7 = \left\{ \begin{array}{c c c} 
     0  & {\rm for} & \dot M < 10^{-7}\msoy \\
     \dot M_{\rm WD} & {\rm for}  & \dot M \geq 10^{-7}\msoy \\
            \end{array} \right.  .
\end{equation}
I.e., $M_7$ gives the amount of mass by which the white dwarf grows at accretion
rates above $10^{-7}\msoy$, including the continued evolution beyond
$M_{\rm WD} = 1.4\mso$.

For System No.~6, the point mass has grown to $1.75\mso$ by the time the
mass transfer rate falls below the critical rate ($\sim 10^{-8}\msoy$).
The quantity $M_7$ in this system is $M_7=0.60\mso$, which means that
were the critical accretion rate as high as $10^{-7}\msoy$, the point mass 
would still have grown to $1.60\mso$. This can be understood from
Fig.~4, which shows the mass transfer rate as function of time
for System No.~6. It can be seen that the mass transfer rate remains
above $10^{-7}\msoy$ for several million years after the potential
supernova explosion. 

Note that this estimate is not fully self-consistent. I.e., were
all the accreted mass lost through nova outbursts as long as
the mass transfer rate were below $10^{-7}\msoy$, the orbital
evolution would differ from that of our model. However,
as the mass loss from the white dwarf would keep the mass ratio
$q:=M_{\rm MS}/M_{\rm WD}$ above one for a longer time, the
orbit would keep shrinking for a longer time,
which would keep the mass transfer rate higher than in our model
(cf. Sect.~3.1).
Thus, the value of $M_7$ which we derive is in fact a lower limit to the
mass which is transfered at rates above $10^{-7}\msoy$.

Comparing Figs.~3 and~4 shows, that the mass transfer rate drops to small
values only several million years after the minimum orbital separation
is achieved. The reason is that at the time of minimum separation
the main sequence star is still more compact
than its thermal equilibrium configuration. I.e., even though the
orbit does not shrink any more, the main sequence star expands
towards its thermal equilibrium radius and drives further mass transfer
thereby.

In Figs.~5 and~6 we show
the long-term evolution of the 
mass transfer rate and of the orbital period
for another system, No.~61, initially consisting of
a low metallicity 1.8$\mso$ main sequence star and a $1\mso$ white dwarf
orbiting with a period of $0.70\,$d. One can see the thermally unstable
mass transfer phase lasting for several million years, beyond which mass
transfer continues only on the nuclear time scale of the main sequence
star, i.e., several $10^9\,$yr. Consequently, the mass transfer rate drops
to some $10^{-10}\msoy$. The system then resembles a Cataclysmic Variable,
evolving on a time scale of $10^9...10^{10}\,$yr. Note that in CVs, this
time scale may become shorter due to angular momentum loss through
magnetic braking (Verbunt \& Zwaan 1981) which is ignored in the present
study. Magnetic braking is not relevant for the 
supernova~Ia progenitor evolution
for two reasons. First, the time scale of the thermally unstable mass transfer 
is only of the order of several million years, which is too short to
allow a significant amount of angular momentum loss through this
mechanism. Second, our main sequence stars do develop convective
envelopes only in the final phase of the thermally unstable mass transfer phase.
I.e., most of the time they have radiative envelopes and thus
supposedly no magnetic wind. For the study of the long term evolution
of those systems which fail to bring the white dwarf to explode as a
supernova, magnetic braking might be relevant. This is, however, beyond the
scope of the present investigation. 

\section{Mass transfer and white dwarf evolution}

\subsection{Mass transfer rates}

\begin{table*}[h!tbp]
\caption{Key properties of interacting main sequence star + white dwarf 
systems with $Z=0.02$.
The columns have the following meanings.
(1) system number,
(2) main sequence star initial mass,
(3) white dwarf initial mass,
(4) initial orbital period,
(5) minimum period,
(6) maximum possible CO-mass (see text),
(7) main sequence star mass when $M_{\rm WD} = 1.4\mso$,
(8) total mass loss due to winds
(9) see Eq.~13,
(10) maximum mass transfer rate
(11) maximum X-ray luminosity of the white dwarf
(12) core hydrogen mass fraction of main sequence star at onset of mass transfer
(13) system age when $M_{\rm WD} = 1.4\mso$
}

\begin{tabular}{c c c c c c c c c c c c c }
\hline
\noalign{\vskip 0.1 truecm}
Nr.& $M_{\rm MS,i}$&$M_{\rm WD,i}$&$P_{\rm i}$&$P_{{\rm min}}$&$M_{{\rm CO}}$&
$M_{\rm MS,f}$&$M_{{\rm wind}}$&$M_7$&$\dot M_{\rm max}$&
$L_{{\rm X}}$&$X_{\rm c}$&$\tau_{\rm SN}$\\
&M$_{\odot}$&M$_{\odot}$&d&d&M$_{\odot}$&M$_{\odot}$&M$_{\odot}$&M$_{\odot}$&
$10^{-7}\msoy$&$10^{38}{{\rm erg\, s}^{-1}}$& &$10^{9}\,$yr\\
\hline
(1)&(2)&(3)&(4)&(5)&(6)&(7)&(8)&(9)&(10)&(11)&(12)&(13)\\
\hline
0&2.4&1.0&1.69&1.49&1.05&-&0.03&0.04&11.0&1.63&0.08& - \\
1&2.3&1.0&0.51&0.27&1.94&1.58&0.28&0.80&7.47&1.57&0.69&0.01\\
2&2.3&1.0&1.74&0.90&1.93&1.47&0.39&0.81&8.68&1.77&0.07&0.79\\
3&2.1&1.0&1.65&1.11&1.85&1.66&0.00&0.72&3.90&1.85&0.08&1.06\\
\noalign{\vskip 0.1 truecm}
4&2.0&1.0&0.69&0.49&1.80&1.56&0.00&0.63&3.75&1.66&0.46&0.77\\
5&2.0&1.0&1.07&0.76&1.77&1.56&0.00&0.63&3.49&1.54&0.24&1.08\\
6&2.0&1.0&1.29&0.91&1.75&1.56&0.00&0.60&3.03&1.34&0.17&1.17\\
7&2.0&1.0&1.63&1.16&1.69&1.55&0.00&0.52&2.35&1.04&0.07&1.26\\
8&2.0&1.0&1.73&1.22&1.70&1.55&0.00&0.52&2.27&1.00&0.03&1.30\\
\noalign{\vskip 0.1 truecm}
9&1.8&1.0&0.54&0.42&1.54&1.35&0.00&0.25&1.45&0.64&0.60&0.70\\
10&1.8&1.0&0.95&0.74&1.53&1.35&0.00&0.28&1.43&0.63&0.28&1.50\\
11&1.8&1.0&1.14&0.88&1.48&1.35&0.00&0.15&1.10&0.48&0.20&1.85\\
12&1.8&1.0&1.22&0.94&1.46&1.35&0.00&0.00&0.97&0.43&0.17&1.89\\
13&1.8&1.0&1.25&0.97&1.45&1.35&0.00&0.00&0.92&0.40&0.15&1.92\\
14&1.8&1.0&1.33&1.02&1.44&-&   0.00&0.00&0.83&0.37&0.11&1.96\\
\\
22&1.74&1.0&0.57&0.47&1.48&1.29&0.00&0.16&1.16&0.51&0.53&1.21\\
23&1.74&1.0&0.63&0.52&1.48&1.29&0.00&0.18&1.23&0.54&0.46&1.70\\
24&1.74&1.0&0.96&0.77&1.44&-&   0.00&0.00&0.94&0.42&0.27&2.01\\
\noalign{\vskip 0.1 truecm}
25&2.1&0.8&0.59&0.28&1.00&-&0.28&0.12&10.5&1.05&0.52& - \\
26&2.1&0.8&0.94&0.42&0.98&-&0.32&0.12&10.3&1.05&0.30& - \\
29&2.1&0.8&1.19&0.43&1.44&0.74&0.72&0.55&10.9&1.05&0.19&0.96\\
30&2.1&0.8&1.23&0.44&1.45&0.75&0.71&0.54&10.6&1.05&0.17&0.98\\
31&2.1&0.8&1.53&0.58&1.49&0.79&0.66&0.58&9.80&1.05&0.09&1.05\\
\noalign{\vskip 0.1 truecm}
32&2.0&0.8&0.60&0.28&1.56&1.05&0.31&0.61&4.49&1.05&0.51&0.66\\
33&2.0&0.8&0.78&0.35&1.54&1.02&0.35&0.61&4.72&1.05&0.37&0.90\\
34&2.0&0.8&1.55&0.76&1.60&1.12&0.24&0.63&4.45&1.05&0.08&1.25\\
35&2.0&0.8&1.79&0.75&1.60&0.66&0.69&0.63&9.08&1.05&0.01&1.31\\
\noalign{\vskip 0.1 truecm}
36&1.9&0.8&0.60&0.33&1.44&1.17&0.09&0.61&3.16&1.05&0.56&0.50\\
37&1.9&0.8&1.66&0.87&1.44&0.93&0.33&0.64&5.05&1.05&0.01&1.61\\
38&1.8&0.8&0.58&0.36&1.50&1.15&0.00&0.50&2.08&0.92&0.53&1.16\\
39&1.8&0.8&1.36&0.83&1.46&1.15&0.00&0.45&1.85&0.82&0.08&2.00\\
\noalign{\vskip 0.1 truecm}
40&1.7&0.8&0.66&0.43&1.38&-&   0.00&0.33&1.40&0.62&0.44& - \\
41&1.7&0.8&0.76&0.50&1.38&-&   0.00&0.34&1.41&0.62&0.37& - \\
42&1.7&0.8&0.81&0.53&1.37&-&   0.00&0.32&1.38&0.61&0.34& - \\
43&1.7&0.8&1.01&0.66&1.35&-&   0.00&0.24&1.21&0.53&0.22& - \\
44&1.7&0.8&1.33&0.87&1.62&1.06&0.00&0.33&1.38&0.61&0.02&2.75\\
45&1.7&0.8&1.50&0.98&1.69&1.06&0.00&0.51&1.99&0.88&0.00&2.76\\
\noalign{\vskip 0.1 truecm}
46&1.6&0.8&0.70&0.47&1.25&-&    0.00&0.00&0.79&0.35&0.42& - \\
47&1.6&0.8&1.11&0.75&1.24&-&    0.00&0.00&0.83&0.37&0.09& - \\
48&1.6&0.8&1.27&0.89&1.51&0.96&0.00&0.37&1.53&0.67&0.00&3.61\\
49&2.0&0.7&0.47&0.25&0.81&-&   0.18&0.10&8.42&1.20&0.64& - \\
50&2.0&0.7&1.49&0.71&0.84&-&   0.22&0.13&8.63&1.21&0.07& - \\
51&1.9&0.7&0.48&0.25&0.87&-&   0.26&0.16&8.73&1.28&0.53& - \\
52&1.9&0.7&1.47&0.55&1.31&-&   0.57&0.48&6.74&0.68&0.07& - \\
53&1.8&0.7&1.37&0.63&1.37&-&   0.27&0.51&3.46&0.68&0.06& - \\
\hline
\end{tabular}
\end{table*}

\begin{table*}[h!tbp]
\caption{List of key properties of computed systems with $Z=0.001$
(cf. Table~2).
}

\begin{tabular}{c c c c c c c c c c c c c }
\hline
\noalign{\vskip 0.1 truecm}
Nr.& $M_{\rm MS,i}$&$M_{\rm WD,i}$&$P_{\rm i}$&$P_{{\rm min}}$&$M_{{\rm CO}}$&
$M_{\rm MS,f}$&$M_{{\rm wind}}$&$M_7$&$\dot M_{\rm max}$&
$L_{{\rm X}}$&$X_{\rm c}$&$\tau_{\rm SN}$\\
&M$_{\odot}$&M$_{\odot}$&d&d&M$_{\odot}$&M$_{\odot}$&M$_{\odot}$&M$_{\odot}$&
$10^{-7}\msoy$&$10^{38}{{\rm erg\, s}^{-1}}$& &$10^{9}\,$yr\\
\hline
(1)&(2)&(3)&(4)&(5)&(6)&(7)&(8)&(9)&(10)&(11)&(12)&(13)\\
\hline
54&1.9&1.0&0.31&0.22&1.77&1.25&0.21&0.68&8.69&1.64&0.61&0.25\\
55&1.9&1.0&0.46&0.32&1.73&1.12&0.34&0.43&11.4&1.70&0.32&0.54\\
56&1.9&1.0&0.53&0.45&1.09&-&   0.06&0.08&11.4&1.68&0.24& - \\
57&1.9&1.0&0.61&0.52&1.08&-&   0.05&0.07&11.3&1.66&0.17& - \\
58&1.9&1.0&0.74&0.65&1.07&-&   0.04&0.06&11.2&1.66&0.06& - \\
59&1.9&1.0&0.79&0.70&1.06&-&   0.04&0.06&11.1&1.65&0.05& - \\
\noalign{\vskip 0.1 truecm}
60&1.8&1.0&0.29&0.22&1.77&1.31&0.04&0.43&5.74&1.69&0.67&0.18\\
61&1.8&1.0&0.70&0.52&1.77&1.07&0.29&0.43&10.3&1.80&0.07&0.79\\
62&1.7&1.0&0.29&0.23&1.68&1.25&0.00&0.42&4.10&1.70&0.68&0.20\\
63&1.7&1.0&0.66&0.53&1.75&1.13&0.13&0.43&7.27&1.80&0.07&0.92\\
64&1.6&1.0&0.29&0.24&1.56&1.15&0.00&0.42&2.85&1.69&0.69&0.20\\
65&1.6&1.0&0.64&0.54&1.70&1.13&0.03&0.43&5.04&1.80&0.08&1.10\\
66&1.5&1.0&0.28&0.25&1.42&-&   0.00&0.24&1.68&0.75&0.74& - \\
67&1.5&1.0&0.61&0.54&1.59&1.06&0.00&0.42&3.36&1.48&0.08&1.32\\
68&1.4&1.0&0.29&0.26&1.30&-&   0.00&0.05&1.47&0.65&0.74& - \\
69&1.4&1.0&0.43&0.40&1.42&-&   0.00&0.27&1.73&0.76&0.27& - \\
70&1.4&1.0&0.58&0.53&1.45&0.96&0.00&0.32&2.09&0.92&0.09&1.60\\
\noalign{\vskip 0.1 truecm}
71&1.8&0.8&0.24&0.17&0.92&-& 0.18&0.12&9.73&1.06&0.67& - \\
72&1.8&0.8&0.68&0.61&0.83&-& 0.03&0.03&8.70&1.05&0.07& - \\
73&1.7&0.8&0.66&0.55&0.85&-& 0.05&0.05&8.96&1.05&0.06& - \\
74&1.6&0.8&0.29&0.19&1.44&-& 0.11&0.53&4.24&1.05&0.67&0.26\\
75&1.6&0.8&0.62&0.40&1.39&-& 0.33&0.50&6.58&1.05&0.07& - \\
76&1.5&0.8&0.28&0.21&1.37&-& 0.00&0.44&2.65&1.05&0.70& - \\
77&1.5&0.8&0.60&0.43&1.40&-& 0.14&0.50&4.42&1.05&0.07& - \\
78&1.4&0.8&0.29&0.23&1.22&-& 0.00&0.22&1.58&0.70&0.73&  - \\
\hline
\end{tabular}
\end{table*}

\begin{figure}[ht]
\begin{centering}
\epsfxsize=0.9\hsize
\epsffile{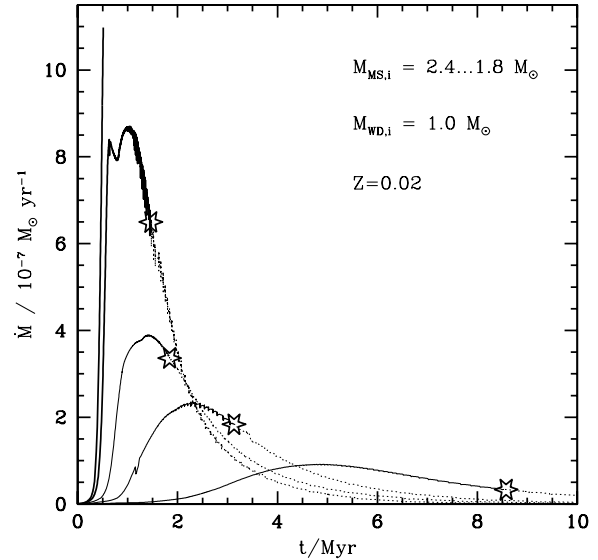}
\caption{Evolution of the mass transfer rate as function of time
for systems No.~0, 2, 3, 7 and 13, which have a metallicity of
Z=0.02 and white dwarf companions with initially
$1\mso$ (see also Fig.~8). System No.0 (leftmost line) is
stopped when the mass transfer rate exceeds the allowed 
upper limit for the wind mass loss rate (cf. Sect.~2.1).
For the other four curves, higher peak mass transfer rates
correspond to larger initial main sequence star masses
(2.3$\mso$, 2.1$\mso$, 2.0$\mso$, and 1.8$\mso$).
The time $t=0$ is defined by the onset of mass transfer.
The star symbol indicates the time
when the white dwarf has reached 1.4$\mso$. Beyond that point,
the graphs are continued as dotted lines.}
\end{centering}
\end{figure}

\begin{figure}[ht]
\begin{centering}
\epsfxsize=0.9\hsize
\epsffile{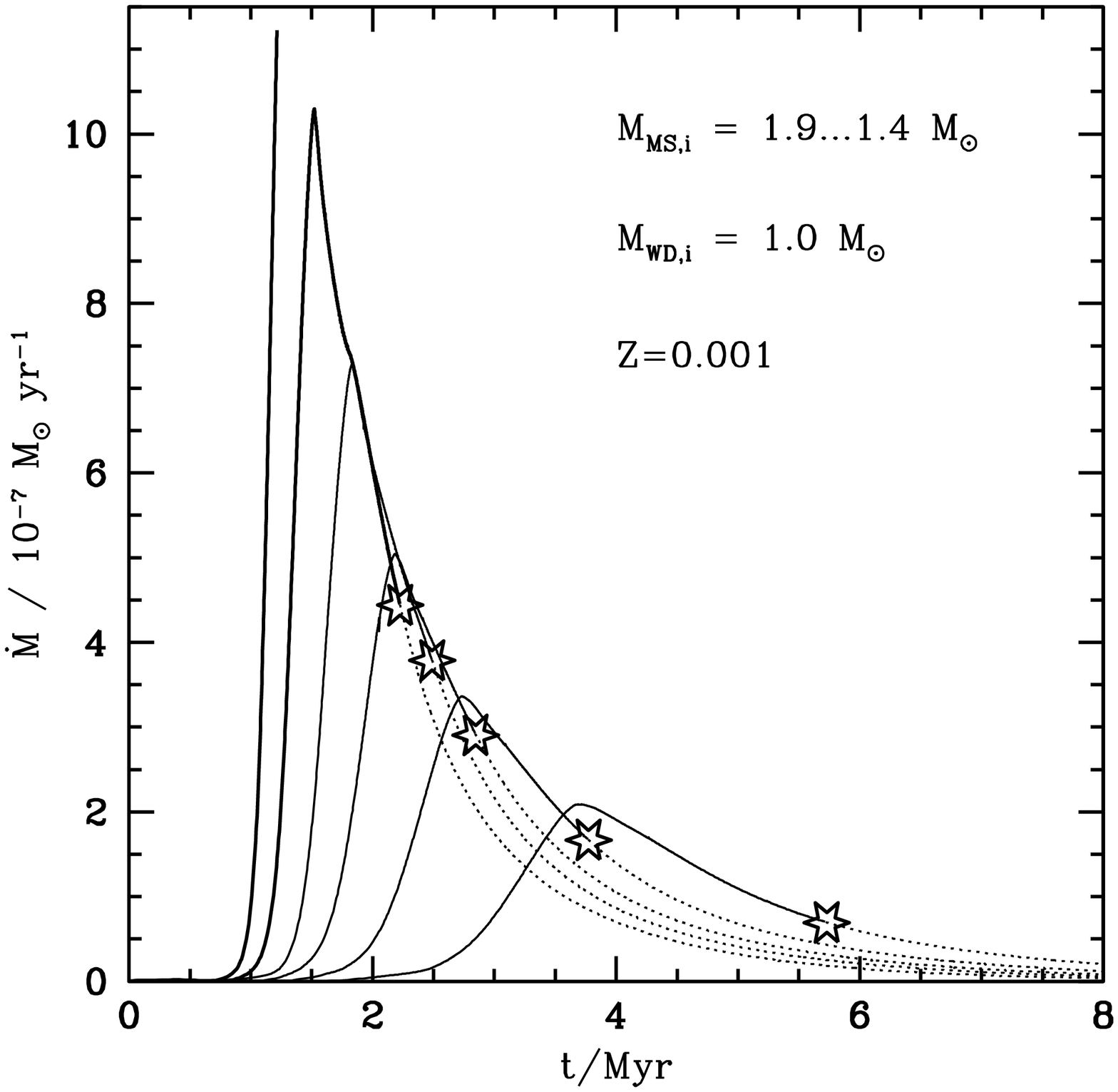}
\caption{Evolution of the mass transfer rate as function of time
for low metallicity systems 
No.~58, 61, 63, 65, 67 and 70. The initial white dwarf mass 
is $1\mso$ for all six cases. Higher peak mass transfer rates
correspond to larger initial main sequence star masses,
from 1.9$\mso$ to 1.4$\mso$ in steps of 0.1$\mso$.
System No.~58 (leftmost line) is stopped when the mass 
transfer rate exceeds the allowed 
upper limit for the wind mass loss rate (cf. Sect.~2.1).
Note that the scale
of the vertical axis is identical to that in Fig.~7, but the
represented initial masses of the main sequence components
is lower.}
\end{centering}
\end{figure}

In this Section, we deal with the mass transfer rates $\dot M$,
and we emphasise that the mass accumulation rate of the white dwarf 
$\dot M_{\rm WD}$ may be smaller than the mass transfer rate if the latter 
is above or below the threshold values defined in Sect.~2.1.
We want investigate the dependence of the mass
transfer rate, and of its time dependence,
on the various initial parameters of our binary systems.
Although Eq.~(13) gives the order of magnitude of the mass transfer rate
during the thermally unstable phase, we will see that 
it fails to reproduce all the physical dependences correctly.

First consider the dependence of the mass transfer rate on the
initial mass of the main sequence
component $M_{\rm MS,i}$. Figures~7 and~8 show the mass transfer rate
as function of time for systems with white dwarf initial masses
of $M_{\rm WD,i} = 1\mso$ and various initial main sequence masses,
for $Z=0.02$ and $Z=0.001$, respectively. For both metallicities,
there is a clear trend to larger maximum mass transfer rates for
more massive main sequence stars caused by                       
the shorter thermal time scale of more massive main sequence stars.
Eq.~(13), with $\tau_{\rm KH}:=G M^2/(2RL)$, does                
reproduce the maximum
mass transfer rates within 30\% for all sequences shown in Fig.~8.
However, it overestimates those of the sequences shown in
Fig.~7 by factors 3...8, larger values corresponding to smaller
initial main sequence star masses.

In all our models, there is a time delay from the
onset of the mass transfer (defined as $t=0$ in Figs.~7 and~8) to the
time when the mass transfer rate has grown sufficiently to allow the
white dwarf mass to grow. This delay is of the order
of the thermal time scale of the main sequence star, i.e. it is
longer for smaller masses. We emphasise that our method to compute the
mass transfer rate (Ritter 1988, see also Braun 1997) allows
its reliable computation 
also for the beginning and the end of the mass transfer evolution. 
Assuming a nova outburst would occur
after the accumulation of $\sim 10^{-5}\mso$ (cf. Prialnik \& Kovetz 1995)
and mass accretion rates of the order of $10^{-8}\msoy$
--- i.e. nova recurrence times of about $10^3\,$yr ---
implies of the order of thousand nova outbursts in our typical 
supernova~Ia progenitors before the white dwarf mass can start growing.

\begin{figure}[ht]
\begin{centering}
\epsfxsize=0.9\hsize
\epsffile{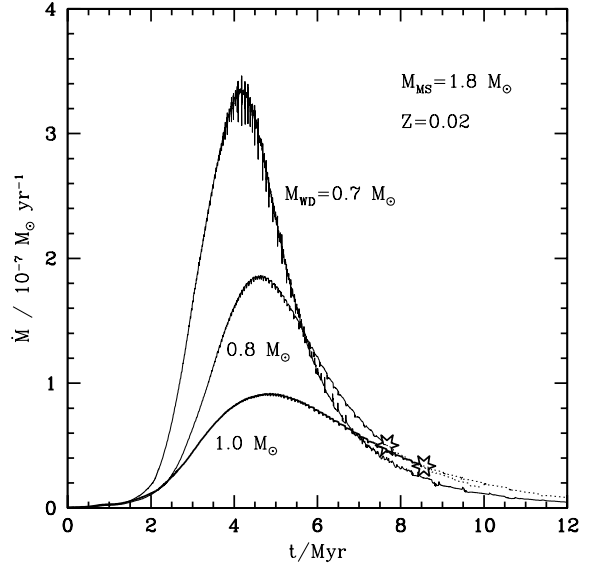}
\caption{Evolution of the mass transfer rate as function of time
for systems No.~13, 39 and 53, i.e. systems with a metallicity
of $Z=0.02$ and with initial main sequence star masses of
$1.8\mso$, but with three different initial white dwarf masses,
as indicated. (Cf. also caption to Fig.~7.)}
\end{centering}
\end{figure}

Fig.~9 shows the mass transfer rate as a function of time for three systems
with the same initial main sequence star mass but with different initial
white dwarf masses. The time delay from
the onset of the mass transfer until the mass transfer rate exceeds
$\sim 10^{-8}\msoy$ and the white dwarf mass can start growing is very
similar for all three systems. The delay
is determined by the thermal time scale of the main sequence star.

Most striking in Fig.~9 is the feature that much larger mass transfer
rates are achieved for smaller white dwarf masses. Although this trend
is also expected from Eq.~(13), the order of magnitude of the effect
seen in Fig.~9 is much larger than what Eq.~(13) predicts.
We find that,
although initially less massive white dwarfs need to accrete more mass 
to reach the Chandrasekhar mass, the best supernova~Ia candidate system 
of those displayed in Fig.~9 may actually be the one with the 
{\em smallest} initial white dwarf mass.
This is so since the same donor star transfers much more mass for
smaller initial white dwarf masses, and that even at higher mass transfer rates.

For otherwise fixed system parameters,
more mass is transfered for smaller initial white dwarf masses since
the minimum orbital separation ---
coincides with the time when $M_{\rm MS}=M_{\rm WD}$
in the conservative case ---
is obtained only after more mass is transferred.
Higher mass transfer rates are achieved
since smaller minimum orbital separations are obtained
for smaller values of $M_{\rm WD,i}$.
For conservative evolution of a given binary system,  
the orbital separation $d$ can be expressed as
\begin{equation}
d = J^2 {M_{\rm MS}+M_{\rm WD} \over G M_{\rm MS}^2 M_{\rm WD}^2} ,
\end{equation}
where
\begin{equation}
J={2\pi d^2 \over P} {M_{\rm MS} M_{\rm WD}\over M_{\rm MS}+M_{\rm WD}}
\end{equation}
is the constant orbital angular momentum.
Therefore, a given initial separation $d_{\rm i}$ relates to the
minimum orbital separation $d_{\rm min}$ as
\begin{equation}
{d_{\rm min}\over d_{\rm i}} = \left( 4 {M_{\rm MS,i} M_{\rm WD,i}
 \over (M_{\rm MS,i} + M_{\rm WD,i})^2 } \right)^2  ,
\end{equation}
and as for fixed system mass the period and separations are related
as $P^2 \propto d^3$ it is
\begin{equation}
{P_{\rm min}\over P_{\rm i}} = \left( 4 {M_{\rm MS,i} M_{\rm WD,i}
 \over (M_{\rm MS,i} + M_{\rm WD,i})^2 } \right)^3  .
\end{equation}
I.e., let us consider two main sequence stars of the same mass, starting to
transfer mass onto their white dwarf companions at the same orbital
separation $d_{\rm i}$. The minimum separation will be smaller for the
binary with the smaller initial white dwarf mass, say System~A.
Since the radius of main sequence stars in the considered mass range
decrease for increasing mass loss rates,
the mass loss rate of the main sequence star in System~A 
--- i.e., its mass transfer
rate --- needs to be larger in order to fit the main sequence star
into a smaller volume. 

The fact that the white dwarf in System~No.~53, which has an initial mass
of $0.7\mso$, does not reach $1.4\mso$ but only $1.37\mso$ is due to
the fact that during the peak of the mass transfer the rate
slightly exceeds the Eddington accretion rate $\dot M_{\rm Edd}$
(which is smaller for smaller white dwarf masses; cf. Sect.~2.1),
and this system loses $0.27\mso$ to a wind. The other two systems
shown in Fig.~9, No.~13 ($M_{\rm WD,i}=1\mso$) and No.~39
($M_{\rm WD,i}=0.8\mso$), which avoid winds, can grow the 
CO-white dwarf to $1.45\mso$ and $1.46\mso$, respectively.
In fact, the system with the largest initial white dwarf mass, 
System~No.~13, is least likely to produce a Type~Ia supernova,
since in this system all mass is transferred at rates below
$10^{-7}\msoy$. I.e., $M_7=0$ in this case, while $M_7=0.45\mso$
for System~No.~39, and $M_7=0.51\mso$ for System~No.~53 (cf. Table~2).

I.e., the effect that systems with smaller initial white dwarf masses are 
better Type~Ia supernova progenitor candidates is only limited
by the smaller upper limits to the white dwarf mass
accumulation rate for smaller white dwarf masses.

\begin{figure}[ht]
\begin{centering}
\epsfxsize=0.9\hsize
\epsffile{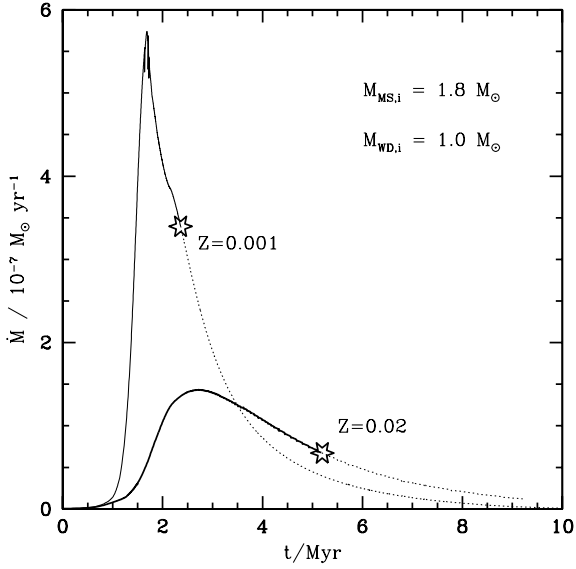}
\caption{Evolution of the mass transfer rate as function of time
for systems No.~9 and 60, which have identical initial
main sequence star and white dwarf masses, but different
metallicities, as indicated. (Cf. also caption to Fig.~7.)}
\end{centering}
\end{figure}

The dependence of the mass transfer rate on the metallicity of the main
sequence star is elucidated in Fig.~10. It shows  two systems
with identical initial main sequence star and initial white dwarf
masses but with different metallicities. Wind mass loss is negligible
in both cases.
The difference in the maximum mass transfer rate of Systems No.~9 and
No.~60 --- which both start with a main sequence star of 1.8$\mso$
and a white dwarf of 1$\mso$ --- is a factor of four. Both systems
have initial periods close to the shortest possible initial period.
This large difference is {\em not} due to different stellar
Kelvin-Helmholtz time scales $\tau_{\rm KH}$. 
Although the low metallicity stars
are more luminous, they are also more compact both effects on
$\tau_{\rm KH}$ almost cancel out (cf. also Table~1). At the onset of
mass transfer, it is $\tau_{\rm KH}=2.6\,$Myr for the main sequence
star in System No.~9, while the corresponding value for System No.~60
is $\tau_{\rm KH}=2.4\,$Myr. This is also reflected in the similarity
of the turn-on times for the mass transfer (cf. Fig.~10).

Low metallicity systems have larger mass transfer rates compared 
to systems with solar abundances (Figs. ~7 and~8).
While the range of mass transfer rates covered
in both figures is the same, Fig.~7 ($Z=0.02$) shows systems with
initial main sequence masses in the range 2.4...1.8$\mso$, while those
in Fig.~8 ($Z=0.001$) are in the range 1.9...1.4$\mso$. 
Based on our detailed models,  Eq.~(13) is valid for low metallicity within
a factor of two. For solar  metallicities, the mass transfer rates are
systematically lower by a factor of~5.
I.e., on average the low metallicity systems have, for the same initial stellar
masses, five times higher mass transfer rates than the systems at $Z=0.02$.

We want to point out that, for the Case~A systems considered in
this work, the maximum mass transfer rate can vary by up to a factor of~2
as function of the initial period (cf. Tables~2 and~3). One would expect
larger mass transfer rates for initially wider systems, since in this case 
the main sequence star is more extended
and more luminous at the onset of the mass transfer, and thus has a
shorter Kelvin-Helmholtz time scales (cf. Table~1). This expectation,
which is also reflected in Eq.~(13), is fulfilled rather well for most
of our low metallicity systems (cf. Table~3). However, at $Z=0.02$ we
find mostly decreasing maximum mass transfer rates for increasing
initial periods and otherwise fixed initial system parameters (cf. Table~2).  
This means that Eq.~(13)
can not be used to predict trends of the mass transfer rate as function
of the initial period or the system metallicity, and shows the
limitations of simplifying approaches to the study of accreting white
dwarfs in binary systems.

\subsection{Evolution of the white dwarf mass}

\begin{figure}[ht]
\begin{centering}
\epsfxsize=0.9\hsize
\epsffile{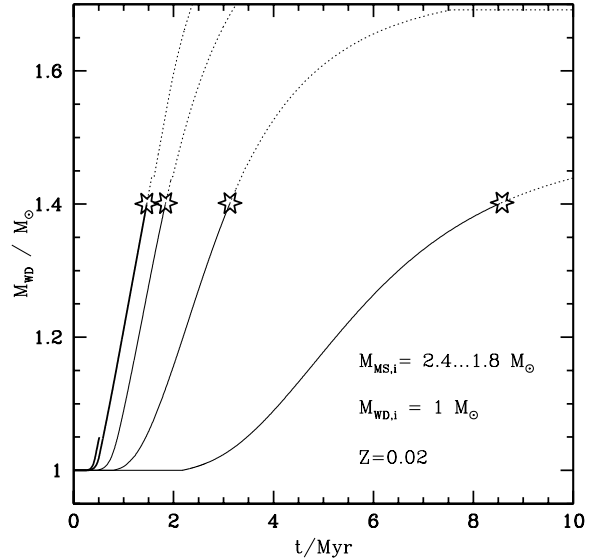}
\caption{Evolution of the white dwarf mass as function of time
for systems No.~0, 2, 3, 7, and 13 (cf. Fig.~7.)}
\end{centering}
\end{figure}

\begin{figure}[ht]
\begin{centering}
\epsfxsize=0.9\hsize
\epsffile{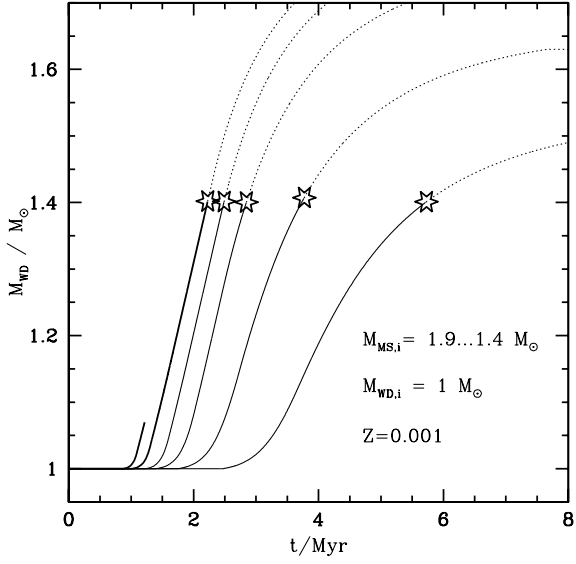}
\caption{Evolution of the white dwarf mass as function of time
for systems No.~58, 61, 63, 65, 67 and 70 (cf. Fig.~8.)}
\end{centering}
\end{figure}

\begin{figure}[ht]
\begin{centering}
\epsfxsize=0.9\hsize
\epsffile{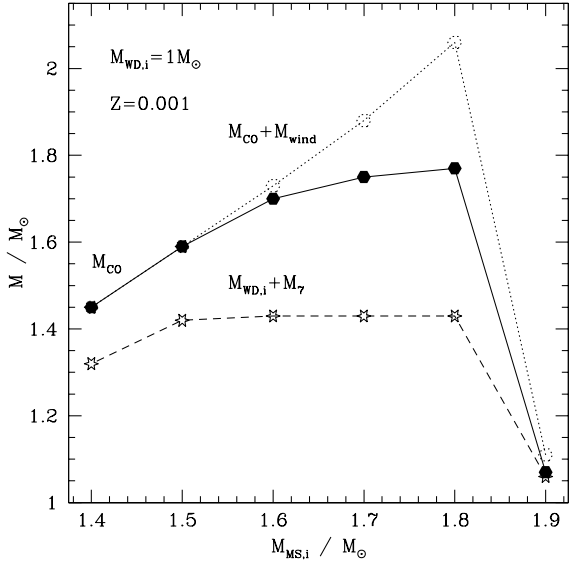}
\caption{Maximum achievable CO-mass $M_{\rm CO}$ (see text; solid line and
solid dots), $M_{\rm CO}$ plus the total mass lost to a wind
(dotted line and dots), and initial white dwarf mass ($1\mso$ for
all systems shown here) plus $M_7$ (cf. Eq.~(14)) --- i.e. the at least
achieved CO-mass in the white dwarf --- for the low metallicity
systems No.~58, 61, 63, 65, 67 and 70 (cf. Figs.~7 and~12, and also 
Figs.~16 and~17).}
\end{centering}
\end{figure}

The dependence of the mass transfer rate on various parameters discussed 
in Sect.~3.1 has important implications for the evolution of the white
dwarfs. Figs.~11 and~12 illustrate the time evolution of the white
dwarf masses for the same systems for which the evolution of the mass
transfer rate $\dot M$ has been displayed in Figs.~7 and~8. We recall
that $\vert \dot M \vert \neq \vert \dot M_{\rm WD} \vert$ due to the
restrictions on the mass accretion rate outlined in Sect.~2. 
I.e., the white dwarf mass can start to grow only 
0.5...3$\,$Myr after the onset of the mass transfer, due to the
occurrence of nova outbursts (cf. also Sect.~3.1).

To demonstrate the effect of the upper and lower critical accretion rates
for the achievable white dwarf masses, we have plotted in Fig.~13
the maximum possible CO-mass in the white dwarf --- ignoring the
possible occurrence of a supernova event at $M_{\rm WD}\simeq 1.4\mso$ --- 
as function of the initial main sequence mass for low metallicity systems
with an initial white dwarf mass of $1\mso$. The sharp drop of the curve
at $M_{\rm MS,i} = 1.9\mso$ is due to the fact that the mass accretion
rate exceeds three times the Eddington accretion rate of the white dwarf
shortly after the onset of the mass transfer in the system with
$M_{\rm MS,i} = 1.9\mso$, which we use as criterion to stop the calculations
(cf. Sect.~2), assuming that the white dwarf would form an extended
hydrogen-rich envelope and the two stars in the system would merge.
 
\begin{figure}[ht]
\begin{centering}
\epsfxsize=0.9\hsize
\epsffile{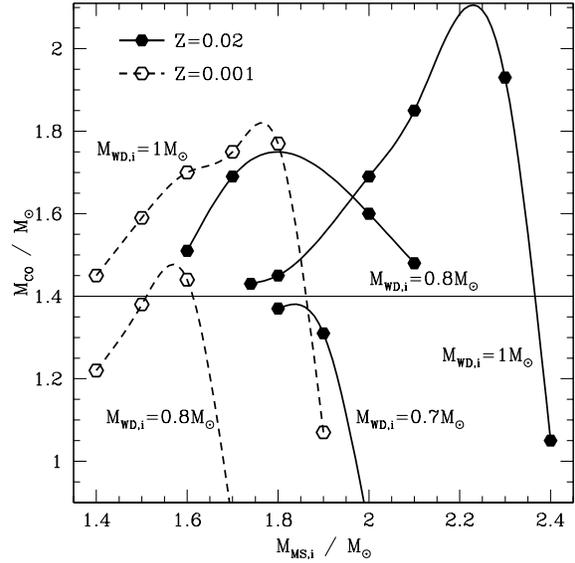}
\caption{Maximum achievable CO-core masses as function of the initial
mass of the main sequence star, for various initial white dwarf masses
and for the two metallicities considered here, as indicated
(cf. also Figs.~13 and ~17).
}
\end{centering}
\end{figure}

\begin{figure}[ht]
\begin{centering}
\epsfxsize=0.9\hsize
\epsffile{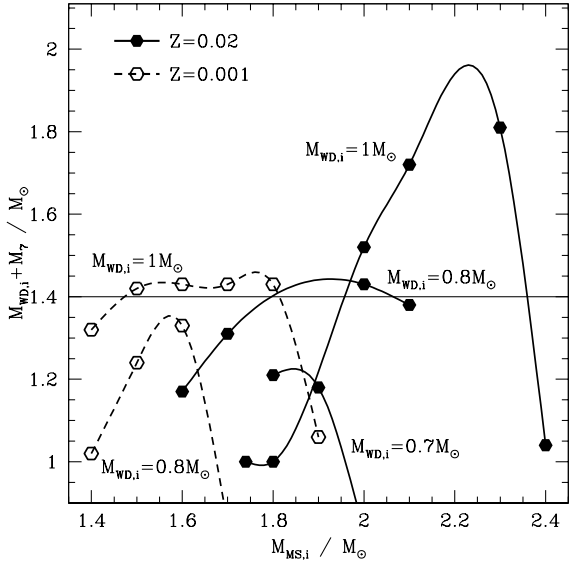}
\caption{CO-core masses achieved if only mass accreted with rates
above $10^{-7}\msoy$ are considered, as function of the initial
mass of the main sequence star, for various initial white dwarf masses
and for the two metallicities considered here, as indicated
(cf. also Figs.~13 and ~16).
}
\end{centering}
\end{figure}

Fig.~13 also shows the sum of $M_{\rm CO}$ and the total amount of mass
lost from the system due to a white dwarf wind (see also Tables~2 and~3).
It indicates that winds, and thus the upper critical accretion rate,
are unimportant for initial main sequence masses 
$M_{\rm MS,i} \simle 1.6\mso$, at $Z=0.001$. 
As discussed in Sect.~2, the wind efficiency is quite uncertain,
particularly at low metallicity.
I.e., Hachisu et al. (1996) and Kobayashi et al. (1998) assume a 
much higher wind efficiency compared to our assumption (Sect.~2),
but assume that the white dwarf winds break down all together at
low metallicity.  A lower wind mass loss rate
might lead to a merging of our systems with $M_{\rm MS,i} \simgr
1.6\mso$ in Fig.~13, rather than to a Type~Ia supernova.
A higher wind mass loss rate, on the other hand, might allow also
white dwarfs in systems with donor star masses larger than 1.8$\mso$ to 
reach the Chandrasekhar mass. We note that, as the
maximum mass transfer rate rises very sharply with increasing initial
main sequence mass (Figs.~7 and~8), the maximum main sequence mass is not very
sensitive to the assumptions on the wind mass loss rate. 
E.g., Li \& van den Heuvel (1997), following Hachisu et al. (1996),
allowed wind mass loss rates of up to $10^{-4}\msoy$, i.e. roughly 
100~times more than in our calculations. This shifts the maximum 
main sequence mass from 2.3$\mso$
in our case to 2.6$\mso$ in their case, 
for Case~A systems with an initial white
dwarf mass of 1$\mso$ and a metallicity of~2\%. 

In order to estimate the effect of a considerably reduced mass
accumulation efficiency of the white dwarf due to weak hydrogen 
shell flashes (cf. Prialnik \& Kovetz 1995) or winds excited
by helium shell flashes (Kato \& Hachisu 1999), 
the lower curve in Fig.~13 shows $M_{\rm WD,i} + M_7$, i.e. the maximum
achievable CO~mass assuming that mass accumulation on the white dwarf is
only possible for $\dot M > 10^{-7}\msoy$ (cf. Eq.~(14)). It shows that
under these assumptions the white dwarfs in the low metallicity
systems with initial main sequence masses of 1.5...1.8$\mso$
would still be able to grow to 1.4$\mso$, although not to significantly
larger values. Furthermore, we note from Fig.~13 (cf. also
Figs.~14 and~15) that a reduction of the limiting mass accretion rate
for mass accumulation by one order of magnitude has little effect
on the upper limit of $M_{\rm MS,i}$ in supernova progenitor systems.

Figs.~14 and~15 show the complete picture of the outcome of our study
for the achievable white dwarf mass as function of the system parameters.
While Fig.~14 gives the optimistic view, i.e. applying the lower
critical accretion rates defined in Sect.~2, Fig.~15 
shows the graphs for $M_{\rm WD,i} + M_7$ rather
than for $M_{\rm CO}$. When considering Figs.~14 and~15, it is important 
to keep in mind that the variation of the initial period of the considered
systems would convert each line in these figures into a band with 
an average width of the order of 0.1$\mso$ (cf. Tables~2 and~3).
These figures allow the following conclusions. 

\begin{enumerate}
\item
The initial
main sequence masses from which Type~Ia supernovae at low metallicity
can be drawn ($\sim 1.45\mso ... 1.85\mso$) are much smaller than at
$Z=0.02$ ($\sim 1.7\mso ... 2.35\mso$). 

\item
The initial masses of the 
white dwarfs required for a Type~Ia supernova are --- on average ---
about 0.2$\mso$ larger at $Z=0.001$ compared to the high metallicity systems.
I.e., note that in Figs.~14 and~15 the curves for $M_{\rm WD,i} = 1\mso$
at $Z=0.001$ are similar to those for $M_{\rm WD,i} = 0.8\mso$ at
$Z=0.02$, only shifted to lower initial main sequence star masses.

\item
Since smaller initial mass ratios $M_{\rm WD,i}/M_{\rm MS,i}$
lead to larger mass transfer rates, 
more mass is transfered in systems with smaller initial white
dwarf masses than for lower initial white dwarf
masses. This effect may give systems with small initial white dwarf masses
--- i.e. perhaps as low as $M_{\rm WD,i}=0.7\mso$ 
--- the possibility to produce a Type~Ia supernova.
\end{enumerate}


\section{Evolution of further system parameters}

\subsection{White dwarf spin}

In Sect.~2.2, we showed that,
within simple approximations, the spin of the accreting white dwarfs
at any given time
depends only on the amount of matter accreted up to that time.
From Eq.~(10) follows that the largest spin at the time of the supernova
explosion is expected in those systems that start out with the least massive
white dwarfs. For initial white dwarf masses of
1.2$\mso$, 1.0$\mso$, 0.8$\mso$, and 0.7$\mso$,
the assumption of homogeneous white dwarfs (i.e., $k^2={2\over 5}$)
leads to ratios of rotational
to critical rotational velocity $\Omega$ of 0.35, 0.67, 0.98, and 1.13,
respectively. 
A more realistic value of $k=0.4$ (Ritter 1985)
results even in $\Omega = 0.85$, 1.68, 2.45, and 2.83.
While these numbers should not be taken
literally --- in particular, values of $\Omega > 1$ are of course not
plausible --- they elucidate the possibility that many of the exploding white
dwarfs in Type~Ia supernovae may be rotating at a speed close to break-up.
According to Figs.~14 and~15 it is not excluded that
white dwarfs with initial masses of $\sim 0.7 \mso$ may contribute
to the Type~Ia supernovae, even at low metallicity.

This point of view is at least partly supported by observational evidence.
While isolated white dwarfs appear to rotate very slowly
($v_{\rm rot} \simle 50\kms$; Heber et al. 1997, Koester et al. 1998)
--- which is also expected from recent single star models with rotation
(Langer et al. 1999) --- those in CVs can be much larger, i.e. up to
$1200\kms$ (Sion 1999). As the white dwarfs in CVs are accreting,
this shows that accreting white dwarfs are in fact spun-up. The fact that
the white dwarfs in CVs are not spinning as rapidly as expected from Eq.~(10)
is interpreted by Livio \& Pringle (1998) as being due to angular momentum
loss in nova explosions which must occur in typical CV systems.

From our simple approach, we expect that by the time the white dwarf
mass gets close to the
Chandrasekhar limit, it may 
rotate with a significant fraction of the
break-up velocity. First polarisation studies of Ia~supernovae seem to indicate
that the degree of polarisation in the supernova spectra is very low
(Wang et al. 1996), which makes a strongly deformed white dwarf as initial
configuration for the explosion rather unlikely.
A confirmation of these results on a solid statistical basis would
imply that either typical initial white dwarf masses are
rather high, or that the white dwarfs can lose angular momentum
during their accretion phase through mechanisms yet to be found.

\subsection{Orbital evolution}

\begin{figure}[t]
\begin{centering}
\epsfxsize=0.9\hsize
\epsffile{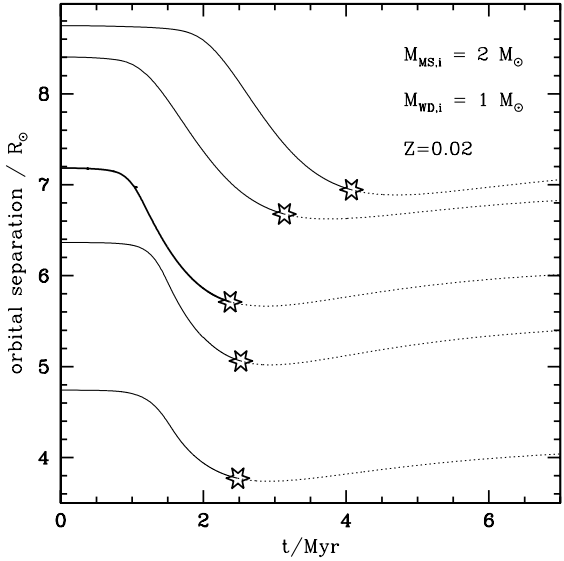}
\caption{Evolution of the orbital separation as function of time
for systems No.~4, 5, 6, 7, and 8. 
The time $t=0$ is defined by the onset of mass transfer.
A star symbol indicates the time
when the white dwarf has reached 1.4$\mso$. Beyond that point,
the graphs are continued as dotted lines.}
\end{centering}
\end{figure}

In all systems the mass transfer rate remains initially on a low level
for about one Kelvin-Helmholtz time scale of the main sequence star.
Nova outbursts are to be expected for the first 0.5...2$\,$Myr of the
mass transfer evolution. This is simulated as a continuous process in our 
calculations, where all transferred mass leaves the system immediately,
carrying the specific orbital angular momentum of the white dwarf.
Both, mass transfer and nova winds at this stage ($M_{\rm MS} > M_{\rm WD}$)
lead to a decrease of the orbital separation. 
The effective shrinkage of the orbit is, however, small
since the amounts of mass involved in the mass transfer and winds
during this phase are small
(cf. Sect.~3). We have not included the possibility of frictional
angular momentum loss during nova outbursts (e.g., Livio \& Pringle 1998),
as this could have been compensated by a slight increase of our 
initial periods which are treated as a free parameter anyway.

Once the mass transfer rate exceeds the critical rates $\dot M_{\rm H}$
and $\dot M_{\rm He}$ (cf. Sect.~2), we assume the mass transfer to be
conservative as long as $\dot M \leq \dot M_{\rm Edd}$ (Eq.~(1)).
In that case, mass and angular momentum conservation lead to a shrinkage
of the orbit as long as $M_{\rm MS} > M_{\rm WD}$, and to a widening
thereafter. Fig.~16 shows the evolution of the orbital separation with
time for five systems with the same initial main sequence star and
white dwarf mass but with different initial periods. As none of these
systems develops a super-Eddington wind, the minimum orbital separations
follow from Eq.~(18) (neglecting the nova winds). 
As Eq.~(18) can be expressed as
\begin{equation}
P_{\rm min} = P_{\rm i} \left( {4 q_i \over (q_i + 1)^2} \right)^3,
\end{equation}
using $q_i = M_{\rm WD,i} / M_{\rm MS,i}$, shorter minimum periods are
achieved for shorter initial periods, and for smaller initial mass ratios
$q_i < 1$. 

Those systems which evolve a super-Eddington wind 
can evolve to considerably shorter periods than
the conservative systems (Eq.~(19)). 
The main reason is that the condition $M_{\rm MS} > M_{\rm WD}$
remains fulfilled for a much longer time than in the conservative case.
E.g., Systems No.~1 and No.~2, which
lose about 0.28$\mso$ and 0.39$\mso$ to a wind, evolve to minimum periods
of 6.5$\,$h and 21.6$\,$h, respectively (cf. Table~1), while their minimum
periods in a conservative evolution would have been 7.5$\,$h and 25.2$\,$h.

\subsection{Evolution of the main sequence star}

\begin{figure}[ht]
\begin{centering}
\epsfxsize=0.9\hsize
\epsffile{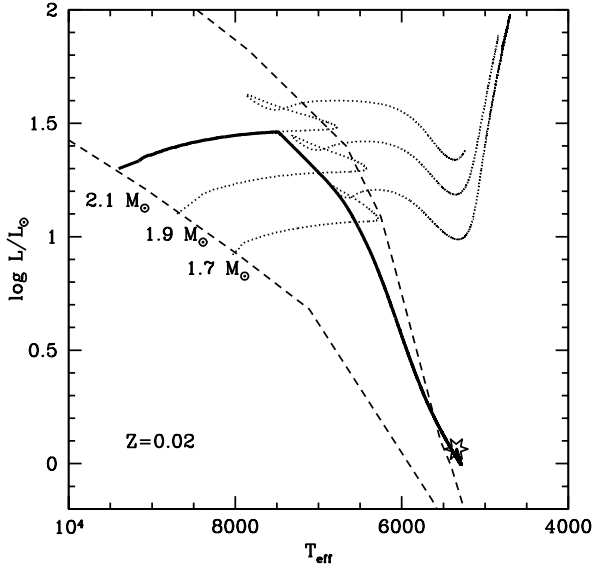}
\caption{Evolutionary track in the HR diagram of the main sequence component
of a 2.1$\mso$+0.8$\mso$ main sequence star~+~white dwarf 
binary with a metallicity of 2\% and
an initial period of 1.23~d (System No.~30),
from zero age until the supernova explosion of the white dwarf 
(thick solid line).
Dotted lines correspond to evolutionary tracks of single stars of
2.1, 1.9, and 1.7$\mso$. The two dashed lines denote the zero age
main sequence (left line) and the terminal age main sequence (corresponding
to core hydrogen exhaustion). The star symbol denotes the time when the white
dwarf reaches 1.4$\mso$.
}
\end{centering}
\end{figure}

A discussion of the properties of the main sequence stars in the
supernova~Ia progenitor systems presented before may be interesting 
for two reasons. First, it might be observable during the accretion
phase, where our systems might appear as supersoft X-ray sources.
Second, the main sequence star may survive the explosion of the
white dwarf and may then serve as an observable witness of the supernova 
progenitor evolution. 

\begin{figure}[ht]
\begin{centering}
\epsfxsize=0.9\hsize
\epsffile{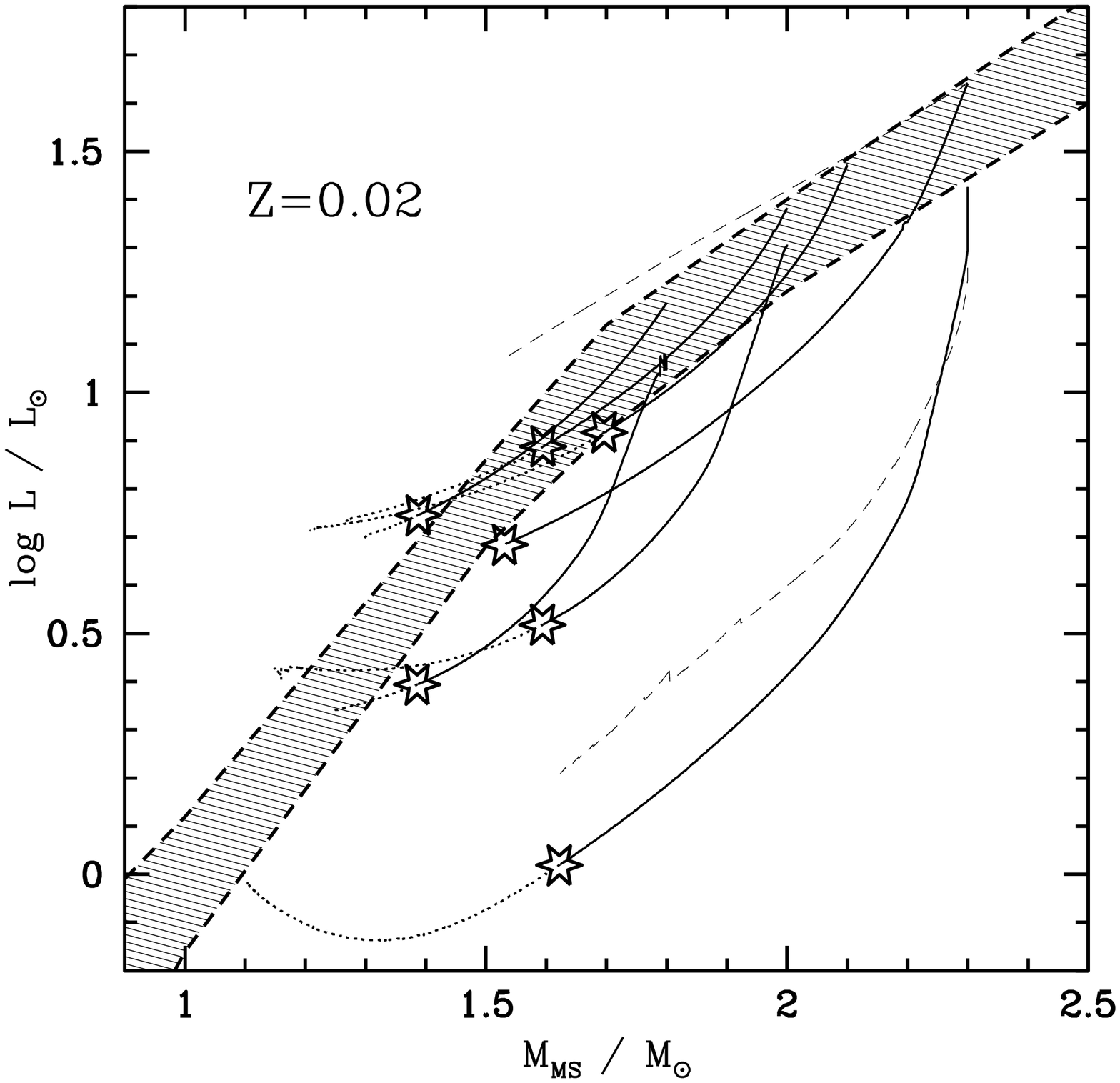}
\caption{Evolutionary tracks of the main sequence stars in
the mass-luminosity diagram starting at the onset of mass transfer, 
for the $Z=Z_{\odot}$ Systems No.~1, 2, 3, 4, 7, 9, 13 
(identifiable by the corresponding initial main sequence star masses,
where the main sequence star of two systems with the same
initial main sequence star mass has a larger initial luminosity
for the system with the larger initial period).
The tracks are shown as continuous lines up to the point when the
white dwarf reaches 1.4$\mso$, and are then continued as dotted lines
(cf. Sect.~3). The shaded band is limited by the dot-dashed lines connecting
the zero age and the terminal age main sequence positions of single stars
in the mass range 0.8...3$\mso$, taken from Schaller et al. (1992). 
The two dashed lines give the evolution of the nuclear luminosity
$L_{\rm nuc} = \int_m \varepsilon_{\rm nuc} dm$ as function of the main
sequence star mass from the onset to the mass transfer to the
time when the white dwarf reaches 1.4$\mso$, for Systems No.~1 (lower
line) and No.~2 (upper line).
}
\end{centering}
\end{figure}

\begin{figure}[ht]
\begin{centering}
\epsfxsize=0.9\hsize
\epsffile{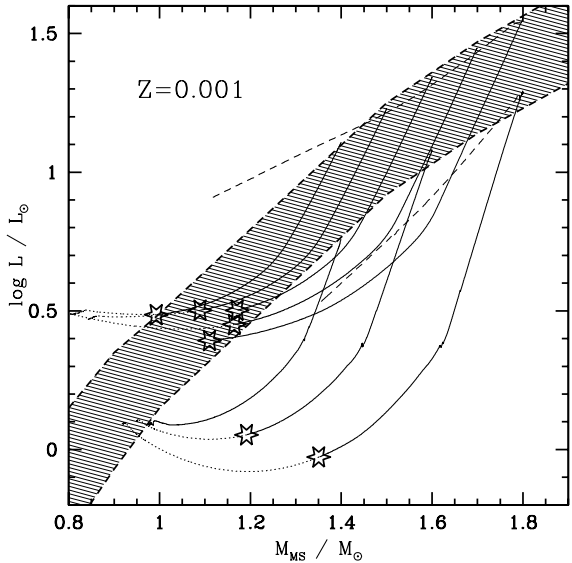}
\caption{Same as Fig.~18, but for our $Z=0.001 Z_{\odot}$ Systems 
No.~60, 61, 63, 64, 65, 67, 68, and~70. The evolution of the
nuclear luminosity is shown for Systems No.~60 (lower line) and
No.~61(upper line).
}
\end{centering}
\end{figure}

\begin{figure}[ht]
\begin{centering}
\epsfxsize=0.9\hsize
\epsffile{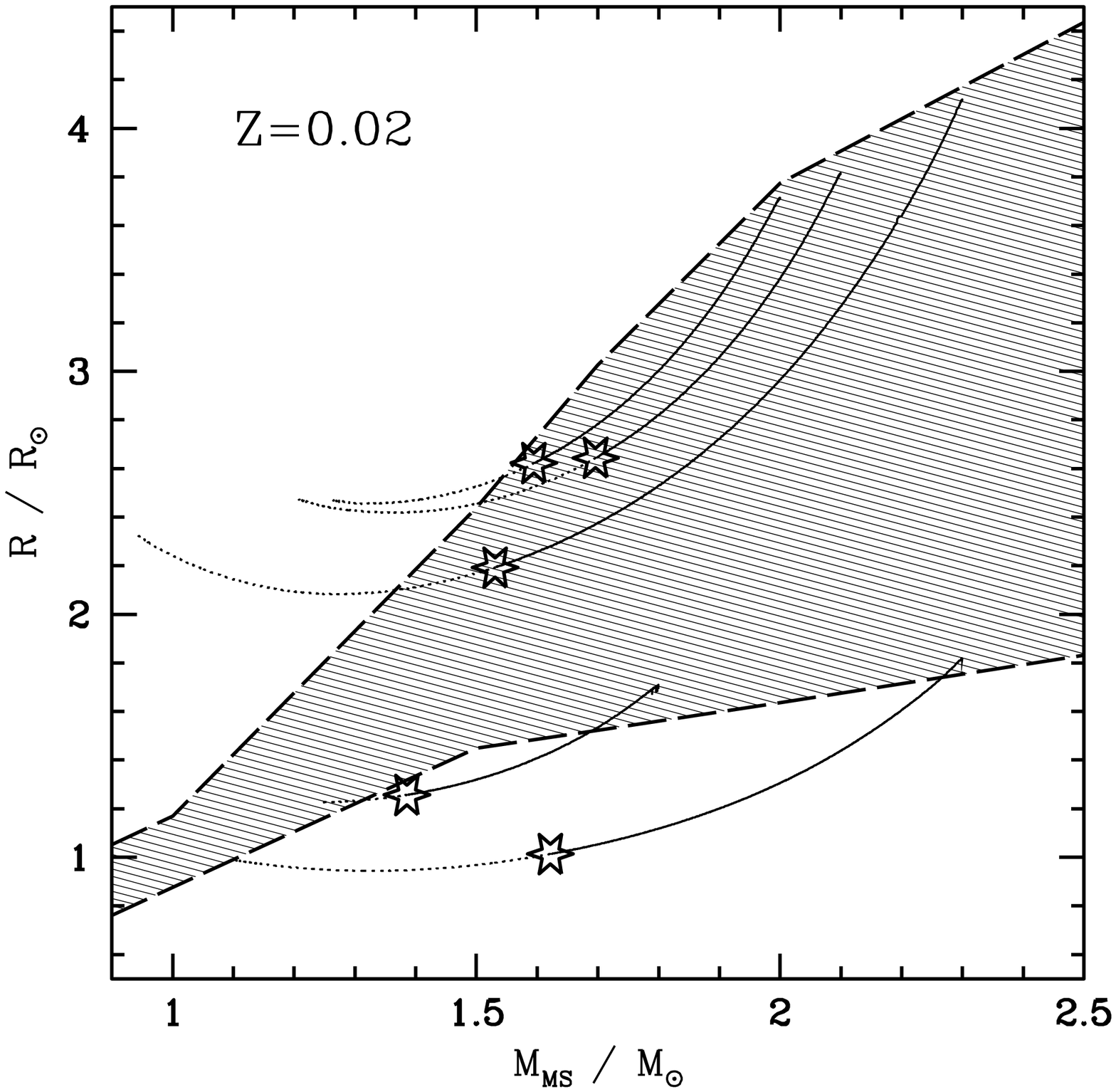}
\caption{Evolutionary tracks of the main sequence stars in
the mass-radius diagram starting at the onset of mass transfer,
for the $Z=Z_{\odot}$ Systems No.~1, 2, 3, 7, 9 
(identifiable by the corresponding initial main sequence star masses,
where the main sequence star of two systems with the same
initial main sequence star mass has a larger initial radius 
for the system with the larger initial period).
The tracks are shown as continuous lines up to the point when the
white dwarf reaches 1.4$\mso$, and are then continued as dotted lines
(cf. Sect.~3). The shaded band is limited by the dot-dashed lines connecting
the zero age and the terminal age main sequence positions of single stars
in the mass range 0.8...3$\mso$, taken from Schaller et al. (1992).
}
\end{centering}
\end{figure}

To elucidate the first point, we have plotted in Fig.~17 the evolutionary track
of the main sequence component of System No.~30 in the HR diagram, in
comparison to normal single stars tracks for comparable initial masses.
It is evident that this star, once the Roche lobe overflow sets in,
reduces its luminosity significantly, roughly by a factor of~$\sim 30$.
However, from Fig.~17 it is not clear which fraction of the
luminosity decrease 
is due to the fact that the stellar mass
of the main sequence star becomes smaller
--- from 2.1$\mso$ to about 0.8$\mso$ at the
time when the white dwarf has reached 1.4$\mso$ ---, and
which fraction is due to the deviation of the star from
global thermal equilibrium during the rapid mass loss phase. 

In order to understand which luminosities the main sequence components of
our systems can achieve in general, and to what extent the reduction of the
luminosity can be understood in terms of the mass reduction, 
we show in Figs.~18 and~19 the mass-luminosity evolution for a
sample of our systems in comparison with the mass-luminosity relation
for normal main sequence stars, for $Z=0.02$ and $Z=0.001$, respectively.
These figures show that in most systems the main sequence
components are significantly underluminous with respect to their mass.

The inspection of Figs.~18 and~19 reveals the following features. First,
the systems with the shorter initial periods, which start out at the
lower boundary of the main sequence band in these figures, evolve
to the zero age main sequence
position corresponding to their final mass once the mass transfer rates
become small and thermal equilibrium is restored. In real binaries the
white dwarf might explode before this happens, but the models shown 
in Figs.~18 and~19 are followed up to the end of the 
thermally unstable mass transfer phase, which corresponds to the endpoint
of the dotted parts of the tracks. 
This result is easily understood, since
when the mass transfer starts these stars are very unevolved, i.e. their 
internal hydrogen and helium distribution is basically homogeneous.
Therefore, when these stars have lost a significant part of their
initial mass, the spatial distribution of their main constituents
(H and He) is still the same as in a zero age main sequence star.
Since the stellar structure adjusts according to the chemical profiles,
and once the mass transfer rate has dropped and the star can relax 
to thermal equilibrium, their properties can not be distinguished
from those of normal main sequence stars of the same mass
(except for trace elements; see below).
 
The deviation of these stars from the zero age main sequence
mass-luminosity relation is due to the thermal imbalance.
Due to the strong mass loss, the outer stellar layers expand which
consumes energy and reduces the stellar luminosity as long as the 
strong mass loss prevails --- strong meaning that the mass loss time scale
is of the same order as the stellar Kelvin-Helmholtz time scale.
This effect by itself can reduce the stellar luminosity by as much
as a factor of~10 (Figs.~18 and~19), i.e., the star can
appear ten times dimmer than expected from the mass-luminosity
relation of single stars on the basis of its actual mass.
The effect is stronger for larger mass loss rates.
We emphasise that this reduction of the stellar luminosity due
to the mass loss induced thermal imbalance comes actually in two
components. One is that due to the strong mass loss, 
the outer stellar layers expand which consumes energy and reduces 
the stellar luminosity below the ``nuclear luminosity'', i.e. the
amount of energy liberated by thermonuclear reactions in the
stellar core. However, as main sequence stars adjust their
nuclear luminosity to the radiative energy loss at the surface
(Kippenhahn \& Weigert 1990), also the nuclear luminosity of the
mass losing main sequence stars is smaller than the nuclear luminosity
of a non-mass losing main sequence star of the same mass and evolutionary
stage (see Figs.~18 and~19). 

For systems with a relatively large initial period, i.e. those where the mass
transfer starts when the main sequence star is close to the terminal age
main sequence (upper borderline of the main sequence band in Figs.~18 and~19),
one phenomenon counterbalances the two dimming effects (i.e. mass reduction
and thermal imbalance). The cores of these stars are, at the onset of the
mass transfer, very helium-rich. Therefore, after the mass transfer
they have a helium-rich core which is significantly more massive 
than a helium core in a single stars of the same stellar mass. 
This makes the stars
overluminous compared to single stars, as can be seen from the dotted
parts of the stellar tracks of the long-period systems shown in
Figs.~18 and~19. Evidently, this effect --- we call it the helium effect ---
is the larger the later the mass transfer starts during the core hydrogen
burning evolution of the main sequence component, and the larger
the total amount of mass lost. 

In summary, we have three effects changing the luminosity of our main
sequence component. First, as its mass is reduced, its luminosity is
reduced according to the mass-luminosity relation of single stars
(the mass effect). Second, the larger the mass loss or mass transfer rate
the more is its luminosity reduced in addition, due
to the thermal imbalance imposed by the mass loss (the mass loss effect).
Third, the helium effect can lead to an increase of the luminosity
for those stars which started out with (and therefore still have) 
relatively large periods. All together, we see from Figs.~18 and~19
that during the thermally unstable mass transfer phase no star can be found
above the single star mass-luminosity band. On the other hand, a significant
fraction of them is found below this band, at luminosities between
1~and 10$\lso$. 

Fig.~20 shows, for the case of the metal-rich stars, that not only the 
luminosities but also the radii of the mass transferring main sequence stars
are significantly smaller than the radii of stars in thermal equilibrium
with the same mass and evolutionary stage. 
Note that there is also a strong dependence of the main sequence star
radii on metallicity (cf. Figs.~1 and~2 in Sect.~2.3).
Both effects may be quite relevant for the
derivation of component masses in supersoft X-ray binaries (cf. Sect.~5.1).

\section{Observable consequences}

\subsection{Supersoft X-ray sources}

\begin{table*}[t]
\caption{Properties of selected systems at the time of the maximum
X-ray luminosity. The columns have the following meanings.
(1) system number (cf. Tables~2 and~3),
(2) main sequence star initial mass,
(3) white dwarf initial mass,
(4) initial period,
(5) mass transfer rate,
(6) mass accumulation rate of white dwarf,
(7) white dwarf luminosity
(8) Eddington luminosity of white dwarf
(9) effective temperature of white dwarf,
(10) white dwarf mass,
(11) main sequence star mass,
(12) main sequence star luminosity,
(13) effective temperature of main sequence star,
(14) orbital period
(15) orbital velocity of main sequence star,
(16) orbital velocity of white dwarf.
}
\begin{tabular}{c c c c c c c c c c c c c c c c}
\hline
\noalign{\vskip 0.1 truecm}
Nr. &$M_{\rm MS,i}$&$M_{\rm WD,i}$&$P_{\rm i}$&$\dot M$&$\dot M_{\rm WD}$&
$L_{X}$&$L_{\rm Edd}$&$T_{\rm WD}$ &$M_{\rm WD}$&
$M_{\rm MS}$&$L_{\rm MS}$&$T_{\rm MS}$ &$P$&$v_{\rm MS}$&$v_{\rm WD}$\\
&&&&$10^{-7}$&$10^{-7}$&$10^{38}$&$10^{38}$&$10^3$&&&&$10^3$&&&\\
&$\mso$&$\mso$&d&{\small $\mso$/yr}&$\mso$/yr&{\small erg/s}&{\small erg/s}&
K&$\mso$&$\mso$&$\lso$&K&d&{\small km/s}&{\small km/s}\\
\hline
(1)&(2)&(3)&(4)&(5)&(6)&(7)&(8)&(9)&(10)&(11)&(12)&(13)&(14)&(15)&(16)\\
\hline
1&2.3&1.0&0.51&7.47&3.87&1.57&1.57&818&1.06&1.91&2.00&6.26&0.32&160&288\\
3&2.1&1.0&1.65&3.90&3.90&1.85&1.85&877&1.25&1.83&10.7&6.17&1.23&117&172\\
4&2.0&1.0&0.69&3.75&3.75&1.66&1.66&837&1.12&1.87&7.03&7.02&0.59&137&229\\
9&1.8&1.0&0.54&1.45&1.45&0.64&1.64&648&1.11&1.67&5.01&7.09&0.47&154&231\\
29&2.1&0.8&1.19&2.39&2.39&1.05&1.51&734&1.02&1.43&1.90&5.73&1.14&114&160\\
34&2.0&0.8&1.55&2.38&2.38&1.05&1.57&739&1.06&1.60&5.56&5.96&1.37&106&160\\
36&1.9&0.8&0.60&2.40&2.38&1.05&1.30&744&0.88&1.82&6.35&7.10&0.54&118&245\\
38&1.8&0.8&0.58&2.08&2.08&0.92&1.30&756&0.88&1.57&5.37&7.01&0.50&130&232\\
40&1.7&0.8&0.66&1.40&1.40&0.62&1.41&636&0.95&1.54&3.98&6.58&0.50&139&225\\
49&2.0&0.7&0.47&8.05&2.67&1.20&1.20&731&0.81&1.71&1.26&6.07&0.25&148&312\\
51&1.9&0.7&0.48&8.60&2.87&1.28&1.28&751&0.87&1.46&0.86&5.73&0.25&167&280\\
\\
54&1.9&1.0&0.31&6.92&3.70&1.64&1.64&833&1.11&1.64&1.80&6.90&0.23&196&290\\
60&1.8&1.0&0.29&5.21&3.81&1.69&1.69&843&1.14&1.60&2.17&7.27&0.24&200&280\\
62&1.7&1.0&0.29&4.07&3.86&1.70&1.70&845&1.15&1.54&2.58&7.54&0.25&127&170\\
64&1.6&1.0&0.29&2.85&2.85&1.25&1.69&783&1.14&1.46&2.65&7.62&0.25&204&261\\
67&1.5&1.0&0.61&3.36&3.36&1.48&1.70&817&1.15&1.35&5.62&7.25&0.54&164&192\\
70&1.4&1.0&0.58&2.09&2.09&0.92&1.67&724&1.13&1.26&5.04&7.19&0.53&167&186\\
71&1.8&0.8&0.24&9.73&2.38&1.06&1.36&725&0.92&1.50&0.65&5.97&0.18&194&314\\

\hline
\end{tabular}
\end{table*}

White dwarfs which accrete hydrogen at such a rate that they can
perform non-explosive hydrogen burning at their surface constitute the
leading model for the persistent supersoft X-ray sources 
(SSSs, Kahabka \& van den Heuvel 1997). Here, we want to compare our
results to observations of SSSs. It is important to keep in mind that
we restricted the parameter space of our models according to the
possibility to obtain a Type~Ia supernova. I.e., although all our models
may be considered as models for SSSs, it is not excluded that
the average SSSs have in fact quite different properties than our models.
We therefore restrict ourselves to investigate three basic observable 
properties, i.e. the X-ray luminosity, the system period, and the
luminosity of the donor star, and rather focus on what the largest
and smallest of these values are rather than considering a typical average.
For this purpose, we have compiled in Table~4 system properties at the
time of the maximum X-ray luminosity --- i.e., at the time of the maximum mass
accumulation rate of the white dwarf, as we assume
$L_X = \varepsilon \dot M_{\rm WD}$ (cf. Sect.~2.1).   

The maximum X-ray luminosity which we can achieve in principle within our
assumptions is that obtained by a Chandrasekhar-mass white dwarf
accreting at its Eddington-rate (cf. Sect.~2.1), i.e. $2.07\, 10^{38}\egs$.
The largest value actually occurring in our models is $1.85\, 10^{38}\egs$
(cf. Table~4). So far, none of the empirical bolometric fluxes derived
from SSSs exceeds this value, although some are quite close to it 
(Kahabka \& van den Heuvel 1997).  

For comparing the periods of our models with those of SSSs, we focus on
the short periods, since also post main sequence donor stars 
can produce SSSs which would occur in longer period systems
(Li \& van den Heuvel 1997, Hachisu et al. 1999, Wellstein et al. 1999b).
At $Z=0.02$, we find periods in the range
1.8...0.25$\,$d (43...6$\,$h), while at $Z=0.001$ periods range from
19...5$\,$h (cf. Tables~2 and~3).
Observed periods in close binary SSSs (Kahabka \& van den Heuvel 1997)
are generally in good agreement with these figures.

Some authors in the literature express the necessity to explain
the shortest periods found in SSSs with alternative scenarios.
E.g., the SMC system 1E0035.4-7230 has a period of 4.1$\,$h
(Schmidtke et al. 1996),
for which van Teeseling \& King (1998) proposed a wind-driven
evolution, with a very low mass main sequence star losing mass induced by the
strong X-ray radiation of the white dwarf. We note that
in particular our low-Z models show periods
as low as 4$\,$h (e.g., System No.~71
in Table~3). Furthermore, according to Eq.~(19) {\em significantly} smaller
periods are achievable for smaller but still plausible initial mass ratios.
I.e., a system starting out with a 2.4$\mso$ main sequence star and
a 0.6$\mso$ white dwarf could reduce is initial period by a factor~4.
Thereby, even periods in the range 2...3$\,$h could be obtained.
Even though such systems might not lead to Type~Ia supernovae,
some of them may still allow for stationary hydrogen burning on the
white dwarf surface for a limited amount of time.

Rappaport et al. (1994), in a population synthesis study of SSSs, 
considered all possible initial masses and periods. 
The shortest periods they find
are of the order of 5$\,$h. They consider only one metallicity
(solar), for which the smallest period we found 
is $\sim 6\,$h. In our low metallicity
systems, we find a minimum of 4$\,$h. Thus, also from
to the results of Rappaport et al. we would expect then minimum
periods as low as 3$\,$h at low metallicity.
Therefore, periods as short as that found in
1E0035.4-7230 may still be explained
within the standard model of thermally unstable mass transfer
studied in the present paper.

Finally, we want to discuss the brightness of the donor stars in 
our models, in relation to the fact that so far none of them
could be observationally identified. In Sect.~4.2 we have seen
that the main sequence stars in our models are, during the mass transfer
phase, significantly underluminous for their actual mass.
In Table~4, we show the properties of the main sequence stars 
at the time of the maximum X-ray luminosity, for selected cases.
Comparing Systems No.~1 and No.~3, we see that the stellar luminosity
is not well correlated with the mass of the main sequence star.
Instead, it is inversely correlated with the mass transfer rate,
i.e. the mass loss rate of the main sequence star.
For systems which have no wind, i.e. for which 
$\dot M = \dot M_{\rm WD}$ and thus $L_X\propto \dot M$,
this means that the brighter the system in X-rays, the dimmer
is the main sequence star. I.e., the 
fact that in the supersoft X-ray sources the X-ray luminosity
is large (otherwise we would not notice them) means that the mass transfer
rate must also be large (cf. Sect.~4.3).
We conclude that the reduction
of the main sequence star luminosity due to the thermal imbalance must
be a large effect in observed supersoft sources of the considered type.
As it can reduce the bolometric luminosity of the main sequence star
by more than one order of magnitude,
it may be quite difficult to observe the main
sequence component in supersoft X-ray binaries.

\subsection{The stellar remnant}

Once the white dwarf has exploded, the main sequence component is likely to
survive, and although some small amounts of mass may be stripped off
by the supernova ejecta and blast wave, (Wheeler et al. 1975, see also
Fryxell \& Arnett 1981, Taam and Fryxell 1984), most stellar properties
of our main sequence components will remain more or less unchanged.
One may hope to identify and observe the 
remaining main sequence component either
in a young Galactic supernova remnant produced by a Type~Ia supernova,
or, if they stick out sufficiently, long after the supernova explosion in the
field.

\begin{table*}[t]
\caption{Ratios of surface abundances of the main sequence star
to initial abundance, for the time of the supernova explosion. The 
abundances of all isotopes of the light elements L1, Be, and~B are zero 
for all models. 
}
\begin{tabular}{c c c c c c c c c c c c c c c}
\hline
\noalign{\vskip 0.1 truecm}
Nr. & $M_{\rm MS,i}$ & $M_{\rm MS,f}$ & $\Delta M$ & $\Delta M_{\rm wind}$ & 
$^3$He & $^4$He & $^{12}$C & $^{13}$C &  $^{14}$N &  
$^{15}$N &  $^{16}$O &  $^{17}$O &  $^{18}$O  & $^{23}$Na \\
~ & M$_{\odot}$ & M$_{\odot}$ & M$_{\odot}$ & M$_{\odot}$ & & & & & & &
& & &\\
\hline
 3 & 2.1 & 1.66 & 0.44 & 0.00 & 6.71 & 1.00 & 1.00 & 1.00 & 1.00 & 1.00 & 1.00 & 1.00 & 1.00 & 1.00 \\
 8 & 2.0 & 1.55 & 0.45 & 0.00 & 7.91 & 1.00 & 1.00 & 1.00 & 1.00 & 1.00 & 1.00 & 1.00 & 1.00 & 1.00 \\
13 & 1.8 & 1.35 & 0.45 & 0.00 & 14.8 & 1.00 & 1.00 & 1.00 & 1.00 & 1.00 & 1.00 & 1.00 & 1.00 & 1.00 \\
29 & 2.1 & 0.74 & 1.36 & 0.72 & 4.33 & 1.00 & 0.08 & 2.22 & 4.61 & 0.06 & 1.00 & 2.92 & 0.02 & 1.06 \\
31 & 2.1 & 0.79 & 1.31 & 0.66 & 6.65 & 1.04 & 0.15 & 3.66 & 4.28 & 0.07 & 1.00 & 2.39 & 0.05 & 1.07 \\
32 & 2.0 & 1.05 & 0.95 & 0.31 & 7.45 & 1.01 & 0.91 & 7.28 & 1.12 & 0.34 & 1.00 & 1.00 & 0.77 & 1.00 \\
35 & 2.0 & 0.66 & 1.34 & 0.69 & 4.10 & 1.07 & 0.08 & 1.91 & 4.60 & 0.06 & 0.99 & 6.27 & 0.03 & 1.07 \\
36 & 1.9 & 1.17 & 0.73 & 0.09 & 16.8 & 1.01 & 0.99 & 1.48 & 1.00 & 0.89 & 1.00 & 1.00 & 0.99 & 1.00 \\
37 & 1.9 & 0.93 & 0.97 & 0.33 & 14.4 & 1.03 & 0.69 & 10.7 & 1.80 & 0.22 & 1.00 & 1.02 & 0.46 & 1.00 \\
38 & 1.8 & 1.15 & 0.65 & 0.00 & 22.6 & 1.00 & 0.99 & 1.10 & 1.00 & 0.97 & 1.00 & 1.00 & 0.99 & 1.00 \\
39 & 1.8 & 1.15 & 0.65 & 0.00 & 26.9 & 1.01 & 0.99 & 1.30 & 1.00 & 0.93 & 1.00 & 1.00 & 0.99 & 1.00 \\
45 & 1.7 & 1.06 & 0.64 & 0.00 & 28.1 & 1.01 & 0.99 & 1.64 & 1.00 & 0.90 & 1.00 & 1.00 & 0.99 & 1.00 \\
\\
54 & 1.9 & 1.25 & 0.65 & 0.21 &  218 & 1.00 & 0.98 & 2.42 & 1.00 & 0.71 & 1.00 & 1.00 & 0.96 & 1.00 \\
55 & 1.9 & 1.12 & 0.78 & 0.34 &  172 & 1.02 & 0.80 & 11.1 & 1.38 & 0.20 & 1.00 & 1.01 & 0.55 & 1.00 \\
60 & 1.8 & 1.31 & 0.49 & 0.04 &  209 & 1.00 & 0.99 & 1.05 & 1.00 & 0.98 & 1.00 & 1.00 & 1.00 & 1.00 \\
61 & 1.8 & 1.07 & 0.73 & 0.29 &  281 & 1.02 & 0.85 & 9.80 & 1.26 & 0.24 & 1.00 & 1.01 & 0.65 & 1.00 \\
62 & 1.7 & 1.25 & 0.45 & 0.00 &  195 & 1.00 & 0.99 & 1.02 & 1.00 & 0.99 & 1.00 & 1.00 & 1.00 & 1.00 \\
63 & 1.7 & 1.23 & 0.47 & 0.13 &  474 & 1.00 & 0.97 & 1.98 & 1.00 & 0.79 & 1.00 & 1.00 & 0.97 & 1.00 \\
64 & 1.6 & 1.15 & 0.45 & 0.00 &  208 & 1.00 & 0.99 & 1.02 & 1.00 & 0.99 & 1.00 & 1.00 & 1.00 & 1.00 \\
65 & 1.6 & 1.13 & 0.47 & 0.03 &  622 & 1.00 & 0.99 & 1.13 & 1.00 & 0.97 & 1.00 & 1.00 & 0.99 & 1.00 \\
67 & 1.5 & 1.06 & 0.44 & 0.00 &  706 & 1.00 & 0.99 & 1.09 & 1.00 & 0.98 & 1.00 & 1.00 & 1.00 & 1.00 \\
\hline
\end{tabular}
\end{table*}

In the first case, the thermal imbalance imposed by the mass transfer
(cf. Sect.~4.4) will still be completely preserved, since the thermal time
scale of the star is of the order of $10^7\,$yr, while any gaseous
supernova remnant would dissolve at least 100~times faster.
The expected luminosities of the main sequence components in a 
supernova~Ia remnant can thus be directly read off Figs.~18 and~19
from the star-symbols, which mark the expected time of the supernova
explosion. They are found to be in the range 1...10$\lso$. When inspecting the
effective temperatures of the main sequence stars at the time of the
supernova, we find them to be systematically 500...1000$\,$K cooler
than single main sequence stars of the same mass and evolutionary state
(Fig.~17; see also column~14 in Table~4).
As the main sequence band has a width of more than 1000$\,$K, this
implies that remnant stars will be located on the main sequence band or 
slightly to the right. At $Z=0.02$, the effective temperatures of the
remnant stars are larger than 5500$\,$K, at $Z=0.001$ larger than 6000$\,$K.
I.e., they would appear as evolved F~or G~type main sequence stars.

Important to unambiguously identify the stellar remnant
of a supernova~Ia progenitor system of the considered type 
is its peculiar surface
chemical composition. These stars have peculiar abundances since
they have lost a major part of their initial mass during the mass  
transfer phase, with the consequence that they uncover matter which has
been sufficiently deep inside the star that thermonuclear reactions 
have occurred.. All main sequence stars in the present study in systems 
which lead to Type~Ia supernovae lose at least $\sim 0.4\mso$, as
the white dwarf needs to achieve the Chandrasekhar mass. However,
in those systems where the white dwarf develops a wind the total mass
loss of the main sequence stars may be considerably larger (cf. Table~5). 

In a normal main sequence star, all isotopes
of the light elements lithium, beryllium and boron are destroyed
in the whole stellar interior except in an outer envelope of
$\sim 0.1\mso$ (or $\sim 0.2\mso$ for boron). Therefore,
the main sequence components
of our systems are, at the time the supernova explosion occurs, all completely
devoid of the light elements. The lack of these elements offers therefore
already an unambiguous way to identify the remnant stars.

In Table~5, we compile other surface abundance anomalies found in our
models. It can be seen, that the isotope $^3$He is overabundant by
a large factor, which is, however, hard if not impossible to verify
observationally at the present time. The same may hold for other isotopic
anomalies, e.g. of $^{13}$C and $^{15}$N. Only for carbon and nitrogen,
we find the possibility of peculiar elemental abundances, i.e., carbon may
be significantly underabundant and nitrogen correspondingly overabundant,  
as the CN-cycle is responsible for this feature.

The supernova explosion can in principle alter the abundances shown in
Table~5 in two ways. It can lead to an additional mass loss of the main
sequence star of the order of 0.1$\mso$...0.2$\mso$
(Wheeler et al. 1975, Fryxell \& Arnett 1981, Taam and Fryxell 1984).
This may lead to a somewhat stronger CN-cycle signature but would not 
change our results qualitatively. It is further not excluded that
it can lead to the deposition
of small amounts of the supernova ejecta on the main sequence star
(Fryxell \& Arnett 1981, Taam and Fryxell 1984). Whether this happens 
or not seems to be unclear at the moment. In any case, the effect might
be some enrichment of the surface composition of the main sequence 
star remnant with the nucleosynthesis products of the supernova,
i.e. in elements between carbon and iron (Thielemann et al. 1986). 

In summary, the light elements, e.g., lithium, and carbon are the most
promising distinguishing chemical characteristics of main sequence type
stellar remnants of Type~Ia supernovae. Another independent characteristic
may be a peculiar radial velocity or proper motion. The main sequence stellar
remnants will at least have a peculiar velocity of the order of their
orbital velocity at the time of the explosion of the white dwarf.
This velocity is in the range 140$\kms$...250$\kms$ for all our systems,
with larger values corresponding to the low metallicity models.
The momentum impacted by the supernova ejecta on the main
sequence star may increase its space velocity up to as much as 
$\sim 500\kms$ (Wheeler et al. 1975). Therefore, any main sequence
type remnant star must have space velocities in a very favourable
range. It is large enough to impose a clearly peculiar kinematic
on the stellar remnant, but it is still much smaller than the velocity
of the supernova ejecta, which implies that the star will remain
for a long time close to the center of the supernova remnant. 
Chemical and kinematic signature together make it in fact a
interesting project to search for a main sequence type stellar remnant in the 
gaseous remnant of the historical Galactic supernova~1006 
(Wellstein et al. 1999b), which is very likely the product of
an exploding white dwarf (Schaefer 1996).

\section{Discussion and conclusions}

We have studied the evolution of close binary systems 
consisting of a main sequence  
star and a white dwarf which are considered as candidates
for progenitors of Type Ia supernovae.
Based on an extended grid of models, we have studied 
the properties of the systems as a 
function of the initial donor star mass, 
initial white dwarf mass, initial period, and chemical composition.
 
Due to our numerical technique (Sect.~2) we obtain, for the first time,
a complete picture of the time dependence of the mass transfer rate in
such systems. We find that the mass transfer rate remains initially
for about one thermal time scale of the main sequence star on a very low
level during which nova outbursts are likely to occur.
Then, the maximum mass transfer rate is rapidly reached. 
We find that most white
dwarfs approach the Chandrasekhar mass during the decline phase of
the mass transfer ($\ddot M_{\rm WD} < 0$; cf. Figs.~7 to~10).
Our results will allow to investigate the effect of this time dependence 
of the mass transfer rate on the lower critical accretion rates 
for stationary nuclear burning on the white dwarf (cf. Sect.~2.1).
This may be important, as Prialnik \& Kovetz (1995) showed that these
threshold values may be smaller for higher white dwarf temperatures.
As for $\ddot M_{\rm WD} < 0$ the white dwarf temperature is expected
to be higher at a given value of $\dot M_{\rm WD}$ compared to the
case of $\ddot M_{\rm WD} =0$, this effect may perhaps 
increase the parameter space of models 
which lead to Chandrasekhar mass white dwarfs.

In contrast to results based on simple estimates of the mass transfer
rate (e.g., Eq.~(13)), we find that the rates increase strongly for
lower initial white dwarf masses (Fig.~9). I.e., even systems with
rather small initial white dwarf masses ($\sim 0.7\mso$) can not be 
excluded to evolve to Type~Ia supernovae. As Chandrasekhar mass
white dwarfs are likely to rotate faster the smaller their initial mass
is, this implies that the white dwarf rotation may be relevant in 
Type~Ia explosions (cf.~Sect.~4.1).

We find that the mass transfer rates in low metallicity systems are,
for the same initial main sequence star and white dwarf masses,
much higher than at solar metallicity. I.e., the initial main sequence
star mass range which results in good Type~Ia supernova progenitor
candidates shifts from $1.6\mso$...$2.3\mso$ to 
$1.4\mso$...$1.8\mso$ (Figs.~14 and~15). 
We note that the exact donor star mass range is uncertain, due to
uncertainties in white dwarf wind mass accumulation efficiencies
(cf. Sects.~2 and 3.2). However, we find that 
at low metallicity, this
range is narrower, and supernova~Ia progenitor systems need to have
white dwarfs which are initially about 0.2$\mso$ more massive
(see also Figs.~14 and~15), leading to a decrease of the 
Type~Ia supernova rate with decreasing metallicity. We note that
this effect differs from the wind inhibition effect proposed by 
Kobayashi et al. (1998).

It is of course tempting to speculate about effects of the metallicity 
dependence of the progenitor evolution on the supernova peak brightness
or decline rate. However, although more realistic calculations
of the white dwarf evolution to the Chandrasekhar mass are now possible,
they need to be performed before definite conclusions can be drawn.
This is so since, although the described effects are likely to introduce
a Z-dependence to the supernova properties, it would interfere
with other such effects as described by H\"oflich et al. (1998),
Dominguez et al. (1999), and Umeda et al. (1999).

It is also hard to disentangle whether possible dependences of
Type~Ia supernova properties on their environment --- as suggested
by Branch et al. (1996) or Wang et al. (1997) --- are due to the
mentioned trends with metallicity or due to different progenitor
types at work. The latter seems more likely considering the
life times of our progenitor models (Tables~2 and~3). Although
our low~Z models invoke lower mass main sequence stars, their predicted
life time is $< 1.5\, 10^9\,$yr, which is roughly similar for our
solar metallicity models. I.e., Type~Ia supernovae in elliptical galaxies,
which may require progenitor life times of $10^{10}\,$yr, can not be
obtained from the type of model presented here, but may rather
require scenarios with low mass red giant (Hachisu et al. 1996) or
CO~white dwarf (Iben \& Tutukov 1984) donor stars.

We emphasise that, nevertheless, models of the considered type
very likely exist in nature, as they correspond to the
close binary supersoft X-ray sources (Kahabka \& van den Heuvel 1997).
The X-ray luminosities, periods, and main sequence star properties
(Sect.~5.1 and Table~4) appear to agree quite well with observed systems.
We also outline a way to test whether such systems can in fact
evolve into Type~Ia supernovae. We make unambiguous predictions
for the chemical and kinematical properties of the stellar remnants
of main sequence star~+~white dwarf systems after the explosion of the
white dwarf (Sect.~5.2), which may be directly tested for the case 
of the historical galactic supernova~1006 (Wellstein et al. 1999b).


\begin{acknowledgements}
  We are very grateful to Hans Ritter and Jochen Greiner 
  for many helpful discussions.
  This work has been supported by the Deutsche Forschungsgemeinschaft
  through grants La~587/15 and~16 and, in part,
  by  NASA Grant LSTA-98-022 and NASA Grant NAG5-3930.
\end{acknowledgements}

\end{document}